\documentclass[a4paper,pra,twocolumn,showpacs,superscriptaddress,groupedaddress]{revtex4}
\usepackage{ae}
\usepackage{physics}
\usepackage[T1]{fontenc}
\usepackage[ansinew]{inputenc}
\usepackage{amsmath}
\usepackage{amssymb}
\usepackage{subcaption}
\usepackage{multirow}
\usepackage{array}
\usepackage[]{graphicx}
\usepackage{wrapfig}
\usepackage{makecell}
\usepackage{xparse}
\usepackage{dcolumn}
\usepackage{color}
\usepackage[colorlinks]{hyperref}
\usepackage{lscape}
\newtheorem{theorem}{Theorem}
\graphicspath{ {./images1/} }

\usepackage{tikz}

\begin{document}
	\title{Strong quantum nonlocality: Unextendible biseparability beyond\\ unextendible product basis}
	\author{Atanu Bhunia}
	\email{atanu.bhunia31@gmail.com}
	\affiliation{Department of Applied Mathematics, University of Calcutta, 92, A.P.C. Road, Kolkata- 700009, India}
	\author{Subrata Bera}
	\email{98subratabera@gmail.com}
	\affiliation{Department of Applied Mathematics, University of Calcutta, 92, A.P.C. Road, Kolkata- 700009, India}
	\author{Indranil Biswas}
	\email{indranilbiswas74@gmail.com}
	\affiliation{Department of Applied Mathematics, University of Calcutta, 92, A.P.C. Road, Kolkata- 700009, India}
	\author{Indrani Chattopadhyay}
	\email{icappmath@caluniv.ac.in}
	\affiliation{Department of Applied Mathematics, University of Calcutta, 92, A.P.C. Road, Kolkata- 700009, India}
	\author{Debasis Sarkar}
	\email{dsarkar1x@gmail.com, dsappmath@caluniv.ac.in}
	\affiliation{Department of Applied Mathematics, University of Calcutta, 92, A.P.C. Road, Kolkata- 700009, India}	
\begin{abstract}
    An unextendible biseparable basis (UBB) is a set of orthogonal pure biseparable states which span a subspace of a given Hilbert space while the complementary subspace contains only genuinely entangled states. These biseparable bases are useful to produce genuinely entangled subspace in multipartite system. Such a subspace could be more beneficial for information theoretic applications if we are able to extract distillable entanglement across every bipartition from each state of this subspace. In this manuscript, we have derived a rule for constructing such a class of UBB which exhibits the phenomenon of strong quantum nonlocality. This result positively answers the open problem raised by Agrawal et al. [Phys. Rev. A 99, 032335 (2019)]; that there exists a UBB which can demonstrate the phenomenon of strong quantum nonlocality in the perspective of local irreducibility paradigm.
\end{abstract}
	\date{\today}
	\maketitle
	\section{INTRODUCTION}
The correlation between quantum entanglement and quantum nonlocality has always been a fundamental area of study in quantum information and foundation theory\cite{EINSTEIN1935}. Apart from ``Bell nonlocality"\cite{Bruner2013,jsbell}, the term ``nonlocality" also gained much attraction in the last two decades with the discovery of quantum ``nonlocality without entanglement"\cite{Bennett1999}. A set of orthogonal product states, which were initially conjectured to be states with classical features has been proven to exhibit purely non-classical phenomenon known as ``nonlocality without entanglement" \cite{Bennett1999}. Data hiding \cite{Terhal2001,Terhal2002,Winter2009,Werner2002}, secret sharing\cite{Markham2008} etc., are some of the applications of this phenomenon.\\
	
Classical information encoded in states of a composite quantum system involving spatially separated subsystems, may not always be decodable under the well known class of operations, known as LOCC (Local operations along with classical communications). Such sets of states are called nonlocal due to their indistinguishable nature under LOCC \cite{Bennett1999,bennett1996,popescu2001,xin2008,Walgate2000,Virmani,Ghosh2001,Groisman,Walgate2002,Divincinzo,Horodecki2003,Fan2004,Ghosh2004,Nathanson2005,Watrous2005,Niset2006,Ye2007,Fan2007,Runyo2007,somsubhro2009,Feng2009,Runyo2010,Yu2012,Yang2013,Zhang2014,somsubhro2009(1),somsubhro2010,yu2014,somsubhro2014,somsubhro2016,Divincenzo2000,Smolin2001,chitamber2014}. To elaborate a little, suppose a state is secretly chosen from a well known set of states of a bipartite system shared between two distant parties, say, Alice and Bob. Their goal is to locally figure out the exact identity of the chosen state. Local quantum state discrimination process plays a prominent role in exploring the restrictions put forward by LOCC \cite{chitamber2014} on quantum systems with spatially separated subsystems. Moreover, contrary to our general intuition, it has been shown that the presence of entanglement in the system in some instances be detrimental to the aforementioned feature of nonlocality of a set of orthogonal states.\\
	
In 1999, Bennet et. al. in their seminal paper \cite{Bennett1999}, first constructed a set of orthogonal product states in $3\otimes 3$ system called unextendible product basis (UPB) that are not perfectly distinguishable under LOCC. The construction was quite striking due to the absence of entanglement and thus providing the fact that entanglement is not an essential feature for the nonlocality of a set of states \cite{Bennett1999,Zhang2015,Wang2015,Chen2015,Yang2015,Zhang2016,Xu2016(2),Zhang2016(1),Xu2016(1),Halder2019strong nonlocality, Halder2019peres set,Xzhang2017,Xu2017,Wang2017,Cohen2008,Zhang2019,somsubhro2018,zhang2018,Halder2018,Yuan2020,Rout2019,bhunia2020,bhunia2022,linchen2016,johnston2013}. In $3\otimes3$ system, five dimensional product basis constitutes a UPB such that no product states lies in the orthogonal complement of the subspace generated by the UPB \cite{Divincinzo,Bennett1999}; i.e., the set of states cannot be extended by adding product states to it while preserving orthogonality. Furthermore, the projector onto the orthogonal complement of UPB is a bound entangled state and hence prescribes a generic rule to construct such states in higher dimensions \cite{Divincenzo2000,Smolin2001}.\\
	
Ever since the discovery of UPB, an interest to study such 'incomplete bases' is on the rise. Recently, Halder et. al.\cite{Halder2019strong nonlocality} came up with the notion of locally irreducible set. It is a set of orthogonal quantum states from which it is not possible to eliminate one or more states by orthogonality preserving local measurements(OPLM). Local irreducibility sufficiently ensures local indistinguishability although the converse is not true. In $3\otimes3\otimes3$ and $4\otimes4\otimes4$ systems, the authors constructed two orthogonal product bases that are locally irreducible in all bipartitions and established the phenomenon of strong quantum nonlocality without entanglement. The authors in \cite{Yuan2020} constructed the strongly nonlocal orthogonal product sets of size $6(d^2-1)$ in $d\otimes d\otimes d$ for $d\geq3$ and a strongly nonlocal orthogonal product basis (SNOPB) in $3\otimes3\otimes3\otimes3$. In a seminal paper \cite{Zhang2019}, authors generalized the definition of strong nonlocality based on the local irreducibility in some multi-partitions and provide some examples in $3\otimes3\otimes3$ and $3\otimes3\otimes3\otimes3$ systems.\\\\

UPBs are very useful to detect entangled states. UPB's complement set does not contain any product state and thus any state form the complementary subspace is essentially entangled. The complement set generates a completely entangled subspace (CES). However, for multiparty system the definition of entangled states is layered. For example, in tripartite system there exists biseparable states which contains entanglement only between some specific pair of parties (for pure states) or combination thereof. The most useful structure of tripartite system is the set form genuinely entangled subspace (GES) which contains all states having entanglement among all of the three parties. Such genuinely entangled states has been proven to be useful in quantum metrology \cite{Hyllus2012,toth2012,Augusiak2016}, quantum key distribution \cite{Nadlinger2022,Xie2021}, quantum secret sharing \cite{Hillery,Walk2021,Chen2015,Long2023}, quantum conference key agreement \cite{Long2023(1),Grasselli2022}, measurement based quantum computation \cite{Briegel2009}, quantum enhanced measurements \cite{Giovannetti2004}, fault tolerant quantum computing \cite{Blanco2021} etc. Thus construction of UBBs are very important in such scenarios for its complement contains only states with genuine entanglement for tripartite systems.\\\\
 
In \cite{agrwal2019} authors first introduce the notion of unextendible biseparable bases (UBB) that provides an adequate method to construct genuinely entangled subspaces (GES). Furthermore, they showed that the GES resulting from the symmetric construction is indeed a bidistillable subspace, i.e., all the states supported on it contain distillable entanglement across every bipartition. In their work, the construction of UBB stems from the structure of UPB and in fact contains UPB as a subset. But this UBB does not exhibit the phenomenon of strong quantum nonlocality since the corresponding UPB does not exhibit the same. Nevertheless in a recent paper \cite{linchen22}, a UPB has been shown to be strongly nonlocal. The corresponding UBB containing it must also be strongly nonlocal due to the inherited UPB substructure. These motivate us to construct a class of UBB without having a UPB as a subset of it and which is strongly nonlocal also.\\\\

In this paper, we have managed to construct such a UBB in $3\otimes3\otimes3$ system and generalized the result for arbitrary higher dimensional cases. The UBB subspace also gives rise to a genuinely entangled subspace(GES). We have proved that this GES is bidistillable, i.e., distillable across every bipartition and thus characterized the GES to some extent. The paper is organized as follows: in Sec.~\ref{A1} necessary definitions and other preliminary concepts are presented. In Sec.~\ref{A2} we construct a $\mathcal{UBB}_\mathcal{II}$ in $C^3\otimes C^3\otimes C^3$ which is strongly nonlocal. In Sec.~\ref{A3} we have succeeded to generalize the result for higher dimensional cases. Finally, the conclusion is drawn in Sec.~\ref{A5} with some open problems for further studies.
	\section{Preliminaries}
	\label{A1}
Every bipartite pure state can be written as $|\psi\rangle=\sum_{i, j} x_{i, j}|i\rangle|j\rangle \in \mathbb{C}^{m} \otimes \mathbb{C}^{n}$, where $|i\rangle$ and $|j\rangle$ are the computational bases of $\mathbb{C}^{m}$ and $\mathbb{C}^{n}$ respectively. There exists a one to one correspondence between the state $|\psi\rangle$ and the $m \times n$ matrix $X=\left(x_{i, j}\right)$. If $\operatorname{rank}(X)=1$, then $|\psi\rangle$ is a product state and if $\operatorname{rank}(X)>1$ then $|\psi\rangle$ is an entangled state. Also, $\left\langle\psi_{1} \mid \psi_{2}\right\rangle=\operatorname{Tr}\left(X_{1}^{\dagger} X_{2}\right)$, where $\left\langle\psi_{1} \mid \psi_{2}\right\rangle$ is the inner product of $\left|\psi_{1}\right\rangle$ and $\left|\psi_{2}\right\rangle$. In a similar manner for a tripartite state $|\phi\rangle=\sum_{i,j,k} y_{i,j,k}|i\rangle|j\rangle|k\rangle \in \mathbb{C}^{m}\otimes\mathbb{C}^{n}\otimes\mathbb{C}^{l}$, where $|i\rangle$, $|j\rangle$ and $|k\rangle$ are the computational bases of $\mathbb{C}^{m}$, $\mathbb{C}^{n}$ and $\mathbb{C}^{l}$ respectively, $|\phi\rangle$ is biseparable if and only if $\operatorname{rank}(Y)=1$ where the matrix $Y=\left(y_{i,j,k}\right)$ written in at least one bipartition and $|\phi\rangle$ is genuinely entangled if and only if $\operatorname{rank}(Y)>1$ in every bipartition. In this section, we will review first some of the definitions which are used throughout the following sections.\\

	{\bfseries Definition 1.} \cite{Halder2018} If all the POVM elements of a measurement structure corresponding to a discrimination task of a given set of states are proportional to the identity matrix, then such a measurement is not useful to extract information for this task and is called $trivial\;measurement$. On the other hand, if not all POVM elements of a measurement are proportional to the identity matrix then the measurement	is said to be a $nontrivial\; measurement$.\\
	
	{\bfseries Definition 2.} \cite{Halder2018} Consider a local measurement to distinguish a fixed set of pairwise orthogonal quantum states. After performing that measurement, if the post measurement states are also pairwise orthogonal to each other then such a measurement is said to be an $orthogonality-preserving\; local\;measurement$(OPLM).\\ 
	
	{\bfseries Definition 3.} \cite{Halder2019strong nonlocality} A set of orthogonal quantum states is $locally\; irreducible$ if it is not possible to eliminate one or more quantum states from the set by nontrivial orthogonality-preserving local measurements.\\
	
	{\bfseries Definition 4.} A set of orthogonal quantum states is $locally\; indistinguishible$  if it is possible to eliminate one or more states from the set by OPLM but not possible to distinguish completely the whole set by nontrivial OPLM.\\
	Therefore it is by definition imply that all locally irreducible states are locally indistinguishable but the converse is not true.\\
	
	A set of pairwise orthogonal product vectors $\left\{|\psi\rangle_{i}\right\}_{i=1}^{n}$ spanning a proper subspace of $\otimes_{j=1}^{m} \mathbb{C}^{d_{j}}$ is called an unextendible product basis (UPB) if its complementary subspace contains no product state \cite{Bennett1999}. Whereas a set of pairwise orthogonal states $\left\{|\psi\rangle_{i}\right\}_{i=1}^{n}$ spanning a proper subspace of $\otimes_{j=1}^{m} \mathbb{C}^{d_{j}}$ is called an unextendible biseparable basis (UBB) if all the states $|\psi\rangle_{i}$ are biseparable and its complementary subspace contains no biseparable state. As all product states are trivially biseparable it is quite possible to extend a set form UPB to UBB but the converse is not always true, i.e, it is not always possible to construct a UPB by reducing some states from a UBB.\\
	
	{\bfseries Definition 5.} A UBB is called $\mathcal{UBB}_\mathcal{I}$ if it contains a UPB as a subset of it and a UBB is called $\mathcal{UBB}_\mathcal{II}$ if it does not contains any UPB as a subset of it.\\
	
	{\bfseries Definition 6.} \cite{Halder2019strong nonlocality} A set of quantum states in tripartite system is said to be $strong \;nonlocal$ if it is locally irreducible in tripartition also locally irreducible in every bipartition.
	
	\section{ construction of strong nonlocal $\mathcal{UBB}_\mathcal{II}$}
	\label{A2}
	Here we provide a rule to construct a complete orthogonal basis of a composite Hilbert space $\mathbb{C}^{n}\otimes\mathbb{C}^{n}\otimes\mathbb{C}^{n}$ by using $n\otimes n$ matrix. For simplicity we provide an example for $\mathbb{C}^{3}\otimes\mathbb{C}^{3}\otimes\mathbb{C}^{3}$.
	Let us choose a $3\otimes3$ matrix $\Delta=\left(\begin{array}{lll}0,0 & 0,1 & 0,2 \\1,0 & 1,1 & 1,2\\2,0 & 2,1 & 2,2\end{array}\right)$ and consider any two conjugate elements $\delta_{mn}$ and $\delta_{nm}$, $(m\neq n\;and\; m,n\in\{0,1,2\})$. Suppose the matrix remaining after removing the row and the column containing the element $\delta_{mn}$ be  $\left(\begin{array}{ll}m',n' & s',t'\\q',r' & o',p' \end{array}\right)$ and the matrix remaining after removing the row and the column containing the element $\delta_{nm}$ be $\left(\begin{array}{ll}m'',n'' & s'',t''\\q'',r'' & o'',p'' \end{array}\right)$. Now we define states\\\\
	 $|\downarrow\kappa^{(i)}_{m,n}\rangle^{\pm}=\ket{n}_{{A_i}}\ket{m'n'\pm o'p'}_{A_{\overline{i+1}}{A_{\overline{i+2}}}}$,\\
	 $|\uparrow\kappa^{(i)}_{m,n}\rangle^{\pm}=\ket{n}_{{A_i}}\ket{q'r'\pm s't'}_{A_{\overline{i+1}}{A_{\overline{i+2}}}}$,\vspace{.06in}\\
	$|\downarrow\kappa^{(i)}_{n,m}\rangle^{\pm}=\ket{m}_{{A_i}}\ket{m''n''\pm o''p''}_{A_{\overline{i+1}}{A_{\overline{i+2}}}}$,\\
	 $|\uparrow\kappa^{(i)}_{n,m}\rangle^{\pm}=\ket{m}_{{A_i}}\ket{q''r''\pm s''t''}_{A_{\overline{i+1}}{A_{\overline{i+2}}}}$,\\
	 where $i\in\{0,1,2\}$ and $\overline{k}$ defines $k\mod3$. As example if $i=2$, $\overline{i+1}=0$ and $\overline{i+2}=1$. So in $\mathbb{C}^{3}\otimes\mathbb{C}^{3}\otimes\mathbb{C}^{3}$ we define a complete orthogonal basis as follows, denote it by $\mathcal{B}_3$.\\
	$|\downarrow \kappa^{(i)}_{m,n}\rangle^{-}$, 
	$|\uparrow \kappa^{(i)}_{m,n}\rangle^{-}$,\\
	$|\updownarrow \kappa^{(i)}_{m,n}\rangle^{-}=|\downarrow \kappa^{(i)}_{m,n}\rangle^{+}-|\uparrow \kappa^{(i)}_{m,n}\rangle^{+}$,\vspace{.06in}\\
	$|\downarrow \kappa^{(i)}_{n,m}\rangle^{-}$, 
	$|\uparrow \kappa^{(i)}_{n,m}\rangle^{-}$,\\
	$|\updownarrow \kappa^{(i)}_{n,m}\rangle^{-}=|\downarrow \kappa^{(i)}_{n,m}\rangle^{+}-|\uparrow \kappa^{(i)}_{n,m}\rangle^{+}$,\\
	$|\kappa^{(i)}_{m,n}\rangle^{\pm}=|\updownarrow\kappa^{(\overline{i+1})}_{m,n}\rangle^{+}\pm|\updownarrow \kappa^{(\overline{i+2})}_{n,m}\rangle^{+}$,\vspace{.06in}\\
	 $\left\{|\kappa_{k}\rangle=|k\rangle_{A_0}|k\rangle_{A_1}|k\rangle_{A_2},k=0,1,2\right\}$.\\
	 
	 Now if we replace the six states $|\kappa^{(i)}_{m,n}\rangle^{\pm}$ in $\mathcal{B}_3$ by $|\kappa^{(i)}_{n,m}\rangle^{\pm}$, we can get another complete basis $\mathcal{B'}_3$ of $\mathbb{C}^{3}\otimes\mathbb{C}^{3}\otimes\mathbb{C}^{3}$. Therefore every conjugate pair defines two complete orthogonal bases for the corresponding composite Hilbert space. Except for the diagonal elements(self conjugate), the matrix $A$ contains three distinct pairs of conjugate elements. Then by this rule, we can define six orthogonal complete bases in $\mathbb{C}^{3}\otimes\mathbb{C}^{3}\otimes\mathbb{C}^{3}$. Now for a particular choice $\{m=0,n=2\}$ we define the complete basis $\mathcal{B}^{0,2}_3$ as follows.\\
	 $|\downarrow \kappa^{(0)}_{0,2}\rangle^{-}=|2\rangle_{A_0}|10-21\rangle_{A_1A_2}$,\;\\
	 $|\downarrow \kappa^{(1)}_{0,2}\rangle^{-}=|2\rangle_{A_1}|10-21\rangle_{A_2A_0}$,\;\\
	 $|\downarrow \kappa^{(2)}_{0,2}\rangle^{-}=|2\rangle_{A_2}|10-21\rangle_{A_0A_1}$,\;\vspace{.06in}\\
	 $|\uparrow \kappa^{(0)}_{0,2}\rangle^{-}=|2\rangle_{A_0}|20-11\rangle_{A_1A_2}$,\\
	 $|\uparrow \kappa^{(1)}_{0,2}\rangle^{-}=|2\rangle_{A_1}|20-11\rangle_{A_2A_0}$,\\
	 $|\uparrow \kappa^{(2)}_{0,2}\rangle^{-}=|2\rangle_{A_2}|20-11\rangle_{A_0A_1}$,\vspace{.06in}\\
	 $|\downarrow \kappa^{(0)}_{2,0}\rangle^{-}=|0\rangle_{A_0}|01-12\rangle_{A_1A_2}$,\;\\
	 $|\downarrow \kappa^{(1)}_{2,0}\rangle^{-}=|0\rangle_{A_1}|01-12\rangle_{A_2A_0}$,\;\\
	 $|\downarrow \kappa^{(2)}_{2,0}\rangle^{-}=|0\rangle_{A_2}|01-12\rangle_{A_0A_1}$,\;\vspace{.06in}\\
	 $|\uparrow \kappa^{(0)}_{2,0}\rangle^{-}=|0\rangle_{A_0}|11-02\rangle_{A_1A_2}$,\\
	 $|\uparrow \kappa^{(1)}_{2,0}\rangle^{-}=|0\rangle_{A_1}|11-02\rangle_{A_2A_0}$,\\
	 $|\uparrow \kappa^{(2)}_{2,0}\rangle^{-}=|0\rangle_{A_2}|11-02\rangle_{A_0A_1}$,\vspace{.06in}\\
	 $|\updownarrow \kappa^{(0)}_{0,2}\rangle^{-}=|\downarrow \kappa^{(0)}_{0,2}\rangle^{+}-|\uparrow  \kappa^{(0)}_{0,2}\rangle^{+}\\
	 \text{\qquad\qquad}=|2\rangle_{A_0}|1-2\rangle_{A_1}|0-1\rangle_{A_2}$,\\
	 $|\updownarrow \kappa^{(1)}_{0,2}\rangle^{-}=|\downarrow \kappa^{(1)}_{0,2}\rangle^{+}-|\uparrow \kappa^{(1)}_{0,2}\rangle^{+}\\
	 \text{\qquad\qquad}=|2\rangle_{A_1}|1-2\rangle_{A_2}|0-1\rangle_{A_0}$,\\
	 $|\updownarrow \kappa^{(2)}_{0,2}\rangle^{-}=|\downarrow \kappa^{(2)}_{0,2}\rangle^{+}-|\uparrow \kappa^{(2)}_{0,2}\rangle^{+}\\
	 \text{\qquad\qquad}=|2\rangle_{A_2}|1-2\rangle_{A_0}|0-1\rangle_{A_1}$,\vspace{.06in}\\
	 $|\updownarrow \kappa^{(0)}_{2,0}\rangle^{-}=|\downarrow \kappa^{(0)}_{2,0}\rangle^{+}-|\uparrow \kappa^{(0)}_{2,0}\rangle^{+}\\
	 \text{\qquad\qquad}=|0\rangle_{A_0}|0-1\rangle_{A_1}|1-2\rangle_{A_2}$,\\
	 $|\updownarrow \kappa^{(1)}_{2,0}\rangle^{-}=|\downarrow \kappa^{(1)}_{2,0}\rangle^{+}-|\uparrow \kappa^{(1)}_{2,0}\rangle^{+}\\
	 \text{\qquad\qquad}=|0\rangle_{A_1}|0-1\rangle_{A_2}|1-2\rangle_{A_0}$,\\
	 $|\updownarrow \kappa^{(2)}_{2,0}\rangle^{-}=|\downarrow \kappa^{(2)}_{2,0}\rangle^{+}-|\uparrow \kappa^{(2)}_{2,0}\rangle^{+}\\
	 \text{\qquad\qquad}=|0\rangle_{A_2}|0-1\rangle_{A_0}|1-2\rangle_{A_1}$,\vspace{.06in}\\
	 $|\kappa^{(0)}_{0,2}\rangle^{\pm}=|\updownarrow \kappa^{(1)}_{0,2}\rangle^{+}\pm|\updownarrow \kappa^{(2)}_{2,0}\rangle^{+}\\
	 \text{\qquad\quad}=|0+1\rangle_{A_0}|21+22\pm(10+20)\rangle_{A_1A_2}$,\;\\
	 $|\kappa^{(1)}_{0,2}\rangle^{\pm}=|\updownarrow \kappa^{(2)}_{0,2}\rangle^{+}\pm|\updownarrow \kappa^{(0)}_{2,0}\rangle^{+}\\
	 \text{\qquad\quad}=|0+1\rangle_{A_1}|21+22\pm(10+20)\rangle_{A_2A_0}$,\;\\
	 $|\kappa^{(2)}_{0,2}\rangle^{\pm}=|\updownarrow \kappa^{(0)}_{0,2}\rangle^{+}\pm|\updownarrow \kappa^{(1)}_{2,0}\rangle^{+}\\
	 \text{\qquad\quad}=|0+1\rangle_{A_2}|21+22\pm(10+20)\rangle_{A_0A_1}$,\;\vspace{.06in}\\
	 $\left\{|\kappa_{k}\rangle=|k\rangle_{A_0}|k\rangle_{A_1}|k\rangle_{A_2},k=0,1,2\right\}$.\\\\
	 Also we define a stopper state\\
	$|S\rangle=(|0\rangle+|1\rangle+|2\rangle)_A(|0\rangle+|1\rangle+|2\rangle)_B(|0\rangle+|1\rangle+|2\rangle)_C$. We claim that the set
	\begin{multline}
    $$
	\mathcal{U}^{0,2}_3=\mathcal{B}_3 \cup\{|S\rangle\} \backslash\Bigl\{\left\{|\kappa^{(0)}_{0,2}\rangle^{+},|\kappa^{(1)}_{0,2}\rangle^{+},|\kappa^{(2)}_{0,2}\rangle^{+}\right\}
	\\
	\cup\left\{|\kappa_k\rangle\right\}_{k=0}^{2}\Bigl\}
	$$
	\label{5}
    \end{multline}
\begin{figure}
\centering
\includegraphics[width=0.52\textwidth]{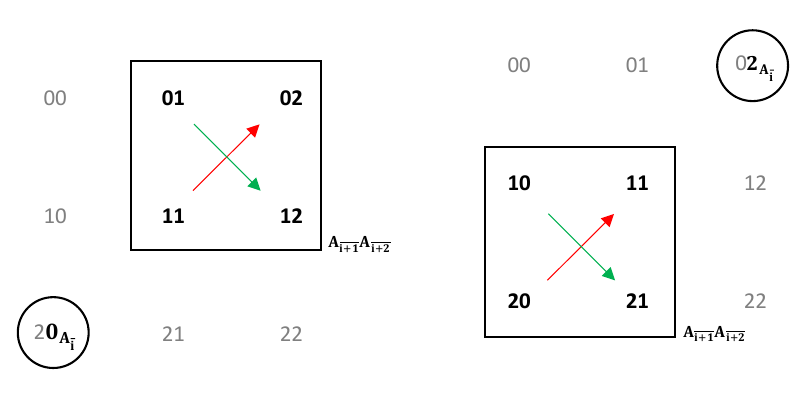}
\caption{ Color outline for the construction of $\mathcal{UBB}_{II}$ in $\mathbb{C}^3\otimes\mathbb{C}^3\otimes\mathbb{C}^3$ as described above. The green arrow relates with $|\downarrow \kappa^{(i)}_{m,n}\rangle^{-}$ whereas the red one relates with $|\uparrow \kappa^{(i)}_{m,n}\rangle^{-}$.}
\label{F1}
\end{figure}
	is a UBB in $\mathbb{C}^{3}\otimes\mathbb{C}^{3}\otimes\mathbb{C}^{3}$. First one can verify that $\mathcal{U}^{0,2}_3$ is an ICOPB. Next the missing states $\left\{\left\{|\kappa^{(0)}_{0,2}\rangle^{+},|\kappa^{(1)}_{0,2}\rangle^{+},|\kappa^{(2)}_{0,2}\rangle^{+}\right\}\cup\left\{|\kappa_k\rangle\right\}_{k=0}^{2}\right\}$ are not orthogonal to $|S\rangle$ but are orthogonal to all states in $\mathcal{U}^{0,2}_3 \backslash$ $\{|S\rangle\}$. Then any state in $\mathcal{H}_{\mathcal{U}^{0,2}_3}^{\perp}$ is a linear combination of at least two of the missing states, and is orthogonal to $|S\rangle$. Assume $|\psi\rangle=a|\kappa^{(0)}_{0,2}\rangle^{+}+b|\kappa^{(1)}_{0,2}\rangle^{+}+c|\kappa^{(2)}_{0,2}\rangle^{+}+h|\kappa_0\rangle+$ $g|\kappa_1\rangle+f|\kappa_2\rangle \in \mathcal{H}_{\mathcal{U}^{0,2}_3}^{\perp}$ is a biseparable state, where at least two coefficients are nonzero. As $|\psi\rangle$ is biseparable state, it is product in at least one bipartition, say $AB|C$. By the correspondence between pure states and matrices, $|S\rangle$(considering in $AB|C$ cut) corresponds to the
	all one matrix $J=\left(\begin{array}{lllllllll}1 & 1 & 1 & 1 & 1 & 1 & 1 & 1 & 1 \\1 & 1 & 1 & 1 & 1 & 1 & 1 & 1 & 1 \\1 & 1 & 1 & 1 & 1 & 1 & 1 & 1 & 1 \end{array}\right)$. Suppose that $|\psi\rangle$ corresponds to a matrix $M=\left(\begin{array}{ccccccccc}h & c & c & a & c & c & a & a & a\\b & b & c & a & g & c & a & a & a\\b & b & c & b & b & c & b & b & f\end{array}\right)$. Since $\operatorname{rank}(M)=1$, we have $a=b=c=h=g=f\neq 0$. However $|\psi\rangle$ (resp. $M$ ) cannot be orthogonal to $|S\rangle$ (resp. $J$ ), and we have a contradiction. We have proved that $\mathcal{U}^{0,2}_3$ is a UBB of size 22 in $\mathbb{C}^{3} \otimes \mathbb{C}^{3}\otimes \mathbb{C}^{3}$.  The color outline for the method of construction is given in FIG.\ref{F1}\\
	As the UBB $\mathcal{U}^{0,2}_3$ do not contain any UPB as a proper subset of it (it forms a $\mathcal{UBB}_\mathcal{II}$) so we cannot assure its nonlocal property (local indistinguishability). 
	Next, we will show that $\mathcal{U}^{0,2}_3$ is nonlocal as well as strongly nonlocal.
	\begin{theorem}
		The set of quantum states (\ref{5}) is locally irreducible in $A_0|A_1|A_2$.
	\end{theorem}
$Proof$: We only need to show any party cannot start a non-trivial OPLM. As we see that the states in (\ref{5}) follow the cyclic property, therefore if any one party (say party $A_0$) goes first and cannot start a nontrivial measurement then $A_1$ and $A_2$ also cannot start nontrivial OPLM. So it is sufficient to prove $A_0$ can only perform the measurement proportional to the identity.\\
	
	Suppose $ E_{\alpha}^{A_0}=M_{\alpha}^{\dag}M_{\alpha} $'s are such measurements that $A_0$ start. As $A_0$'s system is defined in three dimensional Hilbert space $\mathcal{H}^{A_0}$ in $ \{\ket{0}, \ket{1}, \ket{2} \}_{A_0} $ basis, we can write $ E_{\alpha}$ as $3 \cross 3 $ square matrix, as follows:
\begin{equation}
	E_{\alpha}^{A_0}=
	\bordermatrix{
		~ & \bra{0} &\bra{1} & \bra{2} \cr
		\ket{0} & \alpha_{00} & \alpha_{01} & \alpha_{02} \cr
		\ket{1} &\alpha_{10} & \alpha_{11} & \alpha_{12} \cr
		\ket{2} &\alpha_{20} & \alpha_{21} & \alpha_{22}\cr}
\end{equation}
After measurement, all the states either eliminate or remains orthogonal. In both cases $ \mel{\phi}{E_{\alpha}^{A_0} \otimes I_{3}^{A_1} \otimes I_{3}^{A_2}}{\psi} = 0 , \phi \neq \psi ,\phi, \psi \in \mathcal{U}_3^{0,2} $ and for every outcome $ \alpha $.
Then considering the pairs  $ |\downarrow\kappa^{(0)}_{2,0}\rangle^{-}, |\uparrow\kappa^{(2)}_{0,2}\rangle^{-}$ we get 
$$
\begin{array}{l}
	^{-}\left\langle \downarrow\kappa^{(0)}_{2,0}\left|E_{\alpha}^{A_0} \otimes I_{3}^{(1)} \otimes I_{3}^{(2)}\right|\uparrow\kappa^{(2)}_{0,2}\right\rangle^{-} = 0 \text{,}\\
	^{-}\left\langle \uparrow\kappa^{(2)}_{0,2}\left|E_{\alpha}^{A_0} \otimes I_{3}^{(1)} \otimes I_{3}^{(2)}\right|\downarrow\kappa^{(0)}_{2,0}\right\rangle^{-} = 0
\end{array}
$$
	i.e.,
	$$
	\begin{array}{l}
		\mel{0}{E_{\alpha}}{1}_{A_0} \braket{1}{1}_{A_1}\braket{2}{2}_{A_2} = 0  \text{,}\\
		\mel{1}{E_{\alpha}}{0}_{A_0} \braket{1}{1}_{A_1}\braket{2}{2}_{A_2} = 0 
	\end{array}
	$$
	i.e.,
	\begin{equation}
		\alpha_{01}= \alpha_{10}=0
	\end{equation}\\
The complete analysis of the proof is given in the Appendix A.$\blacksquare$\\

In a bipartite quantum system if a set of quantum states is locally irreducible, it means that these states have the strongest nonlocality. But in case of multipartite quantum systems, the presence of entanglement can lead to different strengths of nonlocality among the parties.\\

Now our intention is to show whether the set of states in (\ref{5}) is strongly non-local or not, depending on the local irreducibility of the states in every bipartition.
	\begin{theorem}
		The set of quantum states (\ref{5}) is irreducible in every bipartition.    
	\end{theorem}
$Proof$: Similar to the previous theorem as the set of states given in (\ref{5}) is cyclic in every tripartition, it is also cyclic in every bipartition. So we only need to prove the states are irreducible in $A_0A_1|A_2$, i.e., parties $A_0$ and $A_1$ can apply joint measurement on the subsystem $A_0A_1$.\\
	For that, we rewrite the states in (\ref{5}) in the basis $ \{\ket{0}, \ket{1}, \ket{2}, \ket{3}, \ket{4}, \ket{5}, \ket{6}, \ket{7}, \ket{8} \}_{A_0A_1}$ instead of $ \{\ket{00}, \ket{01}, \ket{10}, \ket{20}, \ket{11}, \ket{02}, \ket{12}, \ket{21}, \ket{22} \}_{A_0A_1} $ respectively, as follows:	
		$$
	\begin{array}{ll}
		|\downarrow \kappa^{(0)}_{0,2}\rangle^{-}= \ket{7}_{\overline{A_0A_1}}\ket{0}_{A_2}-\ket{8}_{\overline{A_0A_1}}\ket{1}_{A_2},\\
		|\downarrow \kappa^{(1)}_{0,2}\rangle^{-} = \ket{51}_{\overline{A_0A_1}A_2} - \ket{62}_{\overline{A_0A_1}A_2},\\
		|\downarrow \kappa^{(2)}_{0,2}\rangle^{-} = \ket{22} - \ket{72},\vspace{.06in}\\
		 |\uparrow \kappa^{(0)}_{0,2}\rangle^{-} = \ket{80} - \ket{71},\\
		|\uparrow \kappa^{(1)}_{0,2}\rangle^{-} = \ket{52} - \ket{61},\\
		 |\uparrow \kappa^{(2)}_{0,2}\rangle^{-} = \ket{32} - \ket{42},\vspace{.06in}\\
		|\downarrow \kappa^{(0)}_{2,0}\rangle^{-} = \ket{01} - \ket{12},\\
		 |\downarrow \kappa^{(1)}_{2,0}\rangle^{-} = \ket{20} - \ket{31},\\
		|\downarrow \kappa^{(2)}_{2,0}\rangle^{-} = \ket{10} - \ket{60},\vspace{.06in}\\
		 |\uparrow \kappa^{(0)}_{2,0}\rangle^{-} = \ket{11} - \ket{02},\\
		|\uparrow \kappa^{(1)}_{2,0}\rangle^{-} = \ket{21} - \ket{30},\\
		 |\uparrow \kappa^{(2)}_{2,0}\rangle^{-} = \ket{40} - \ket{50},\vspace{.06in}\\
		|\updownarrow \kappa^{(0)}_{0,2}\rangle^{-} = \ket{7-8}\ket{0-1},\\
		 |\updownarrow \kappa^{(1)}_{0,2}\rangle^{-} = \ket{5-6}\ket{1-2},\\
		|\updownarrow \kappa^{(2)}_{0,2}\rangle^{-}=\ket{2-3-4+7}\ket{2},\vspace{.06in}\\
		 |\updownarrow \kappa^{(0)}_{2,0}\rangle^{-} = \ket{0-1}\ket{1-2},\\
		|\updownarrow \kappa^{(1)}_{2,0}\rangle^{-}=\ket{2-3}\ket{0-1},\\
		 |\updownarrow \kappa^{(2)}_{2,0}\rangle^{-}=\ket{1-4-5+6}\ket{0},\vspace{.06in}\\
		|\kappa^{(0)}_{0,2}\rangle^{-} =\ket{5+6}\ket{1+2}-\ket{1+4+5+6}\ket{0},\\
		|\kappa^{(1)}_{0,2}\rangle^{-} = \ket{2+3+4+7}\ket{2}-\ket{0+1}\ket{1+2},\\
		|\kappa^{(2)}_{0,2}\rangle^{-} =\ket{7+8-2-3}\ket{0+1},\vspace{.06in}\\
		\ket{S} = \ket{0+1+2+3+4+5+6+7+8}\ket{0+1+2}.
	\end{array}
	$$
	The proof is quite similar to the previous one. $A_0A_1$ starts an OPLM $ E_{\alpha}^{A_0A_1}=M_{\alpha}^{\dag}M_{\alpha} $ which is nothing but a square matrix of order 9.
	\begin{equation}
		E_{\alpha}^{A_0A_1}=
		\bordermatrix{
			~ & \bra{0} &\bra{1} & \cdots & \bra{8} \cr
			\ket{0} & \alpha_{00} & \alpha_{01} & \cdots & \alpha_{08} \cr
			\ket{1} &\alpha_{10} & \alpha_{11} & \cdots & \alpha_{18} \cr
			\vdots & \vdots & \vdots & \ddots & \vdots \cr
			\ket{8} &\alpha_{80} & \alpha_{81} & \cdots & \alpha_{88}\cr}
	\end{equation}
	As we know $ M_{\alpha} \otimes I_{A_2}\ket{\phi} $'s for $ \ket{\phi} \in \mathcal{U} $  are mutually orthogonal, for every order pairs $\{\ket{\psi},\ket{\phi}\}, \ket{\phi} \neq \ket{\psi} \in \mathcal{U} $ and for every outcome $ \alpha $, $ \mel{\psi}{E_{\alpha}^{A_0A_1} \otimes I_{3}^{(2)}}{\phi} = 0$.	
	Now considering the order pairs $ \{\ket{\psi},\ket{\phi}\} $  for $\ket{\psi} \in \{	|\downarrow \kappa^{(1)}_{2,0}\rangle^{-},	|\uparrow \kappa^{(1)}_{2,0}\rangle^{-},	|\downarrow \kappa^{(0)}_{0,2}\rangle^{-},	|\uparrow \kappa^{(0)}_{0,2}\rangle^{-}\}$ and $\ket{\phi}\in\{|\downarrow\kappa^{(1)}_{0,2}\rangle^{-},	|\uparrow \kappa^{(1)}_{0,2}\rangle^{-},	|\downarrow \kappa^{(0)}_{2,0}\rangle^{-},	|\uparrow \kappa^{(0)}_{2,0}\rangle^{-}\}$, we get
	$ \alpha_{ij}=0$ (and hence, $ \alpha_{ji}=0$ ) for $ i=3,2,8,7 $ and $j=5,6,0,1 $ respectively. 
	\begin{equation}
		\therefore,~~ E_{\alpha}^{A_0A_1}=
		\begin{bmatrix}
			\alpha_{00} & \alpha_{01} & 0 & 0 & \alpha_{04} & \alpha_{05} & \alpha_{06} & 0 & 0 \\
			\alpha_{10} & \alpha_{11} & 0 & 0 & \alpha_{14} & \alpha_{15} & \alpha_{16} & 0 & 0 \\
			0 & 0 & \alpha_{22} & \alpha_{23} & \alpha_{24} & 0 & 0 & \alpha_{27} & \alpha_{28} \\
			0 & 0 & \alpha_{32} & \alpha_{33} & \alpha_{34} & 0 & 0 & \alpha_{37} &  \alpha_{38} \\
			\alpha_{40} & \alpha_{41} & \alpha_{42} & \alpha_{43} & \alpha_{44} & \alpha_{45} & \alpha_{46} & \alpha_{47} & \alpha_{48} \\
			\alpha_{50} & \alpha_{51} & 0 & 0 & \alpha_{54} & \alpha_{55} & \alpha_{56} & 0 & 0 \\
			\alpha_{60} & \alpha_{61} & 0 & 0 & \alpha_{64} & \alpha_{65} & \alpha_{66} & 0 & 0 \\
			0 & 0 & \alpha_{72} & \alpha_{73} & \alpha_{74} & 0 & 0 & \alpha_{77} & \alpha_{78} \\
			0 & 0 & \alpha_{82} & \alpha_{83} & \alpha_{84} & 0 & 0 & \alpha_{87} & \alpha_{88} 
		\end{bmatrix}
	\end{equation}
 The complete analysis of the proof is given in Appendix B.$\blacksquare$\\
 	A UBB can be constructed in different ways, our construction stems without multipartite unextendible product bases (UPB). The unextendibility feature of a multipartite UPB is generally not preserved under different spatial configurations, i.e., when you change the way of distribution of quantum states among different parties, the unextendibility property may no longer hold.  Such a multipartite UPB can
 be converted into a complete orthogonal base by allowing entanglement among a subset of parties only. When unextendibility is guaranteed across various arrangements, it leads to different classes of entangled subspaces and the most constrained one among these is the Genuine Entangled Subspace (GES).\\
 
 Using the structural elegance we generalize the above constructions in $\left(\mathbb{C}^n\right)^{\otimes 3}$, with $n \geqslant 3$. In the next section we provide the explicit construction for $n=5$ and then provide the generalization for arbitrary dimension.
 \section{$\text {CONSTRUCTIONS IN }\left(\mathbb{C}^5\right)^{\otimes 3}$}
 \label{A3}
	In this section, we consider the $\mathcal{UBB}_\mathcal{II}$ in higher dimensional cases. As a similar manner we can construct the $\mathcal{UBB}_\mathcal{II}$ in $\mathbb{C}^5\otimes\mathbb{C}^5\otimes\mathbb{C}^5$.  Now for a particular choice $m,n\in\{0,4\}$ we define the complete basis $\mathcal{B}^{0,4}_5$ as follows.\\
	(For $i=0,1,2$)\\
	$|\downarrow\phi^{(i)}_{0,4}\rangle^{-}=|4\rangle_{A_i}(|10\rangle-|21\rangle)_{A_{\overline{i+1}}A_{\overline{{i+2}}}}$,\\
	$|\uparrow\phi^{(i)}_{0,4}\rangle^{-}=|4\rangle_{A_i}(|20\rangle-|11\rangle)_{A_{\overline{i+1}}A_{\overline{{i+2}}}}$,\\ 
	$|\updownarrow\phi^{(i)}_{0,4}\rangle^{-}=|\downarrow\phi^{(i)}_{0,4}\rangle^{+}-|\uparrow\phi^{(i)}_{0,4}\rangle^{+}$,\vspace{.06in}\\
	$|\downarrow\psi^{(i)}_{0,4}\rangle^{-}=|4\rangle_{A_i}(|12\rangle-|23\rangle)_{A_{\overline{i+1}}A_{\overline{{i+2}}}}$,\\
	$|\uparrow\psi^{(i)}_{0,4}\rangle^{-}=|4\rangle_{A_i}(|22\rangle-|13\rangle)_{A_{\overline{i+1}}A_{\overline{{i+2}}}}$,\\
	$|\updownarrow\psi^{(i)}_{0,4}\rangle^{-}=|\downarrow\psi^{(i)}_{0,4}\rangle^{+}-|\uparrow\psi^{(i)}_{0,4}\rangle^{+}$,\vspace{.06in}\\
	$|\downarrow\xi^{(i)}_{0,4}\rangle^{-}=|4\rangle_{A_i}(|30\rangle-|41\rangle)_{A_{\overline{i+1}}A_{\overline{{i+2}}}}$,\\
	$|\uparrow\xi^{(i)}_{0,4}\rangle^{-}=|4\rangle_{A_i}(|40\rangle-|31\rangle)_{A_{\overline{i+1}}A_{\overline{{i+2}}}}$,\\
	$|\updownarrow\xi^{(i)}_{0,4}\rangle^{-}=|\downarrow\xi^{(i)}_{0,4}\rangle^{+}-|\uparrow\xi^{(i)}_{0,4}\rangle^{+}$,\vspace{.06in}\\
	$|\downarrow\eta^{(i)}_{0,4}\rangle^{-}=|4\rangle_{A_i}(|32\rangle-|43\rangle)_{A_{\overline{i+1}}A_{\overline{{i+2}}}}$,\\
	$|\uparrow\eta^{(i)}_{0,4}\rangle^{-}=|4\rangle_{A_i}(|42\rangle-|33\rangle)_{A_{\overline{i+1}}A_{\overline{{i+2}}}}$,\\
	$|\updownarrow\eta^{(i)}_{0,4}\rangle^{-}=|\downarrow\eta^{(i)}_{0,4}\rangle^{+}-|\uparrow\eta^{(i)}_{0,4}\rangle^{+}$,\vspace{.06in}\\
	$|\downarrow\kappa^{(i)}_{0,4}\rangle^{-}=|\updownarrow\phi^{(i)}_{0,4}\rangle^{+}-|\updownarrow\eta^{(i)}_{0,4}\rangle^{+}$,\\
	$|\uparrow\kappa^{(i)}_{0,4}\rangle^{-}=|\updownarrow\xi^{(i)}_{0,4}\rangle^{+}-|\updownarrow\psi^{(i)}_{0,4}\rangle^{+}$,\\
	$|\updownarrow\kappa^{(i)}_{0,4}\rangle^{-}=|\downarrow\kappa^{(i)}_{0,4}\rangle^{+}-|\uparrow\kappa^{(i)}_{0,4}\rangle^{+}$,\vspace{.15in}\\
	$|\downarrow\phi^{(i)}_{4,0}\rangle^{-}=|0\rangle_{A_i}(|01\rangle-|12\rangle)_{A_{\overline{i+1}}A_{\overline{{i+2}}}}$,\\
    $|\uparrow\phi^{(i)}_{4,0}\rangle^{-}=|0\rangle_{A_i}(|11\rangle-|02\rangle)_{A_{\overline{i+1}}A_{\overline{{i+2}}}}$,\\ 
    $|\updownarrow\phi^{(i)}_{4,0}\rangle^{-}=|\downarrow\phi^{(i)}_{4,0}\rangle^{+}-|\uparrow\phi^{(i)}_{4,0}\rangle^{+}$,\vspace{.06in}\\
    $|\downarrow\xi^{(i)}_{4,0}\rangle^{-}=|0\rangle_{A_i}(|03\rangle-|14\rangle)_{A_{\overline{i+1}}A_{\overline{{i+2}}}}$,\\
    $|\uparrow\xi^{(i)}_{4,0}\rangle^{-}=|0\rangle_{A_i}(|13\rangle-|04\rangle)_{A_{\overline{i+1}}A_{\overline{{i+2}}}}$,\\
     $|\updownarrow\xi^{(i)}_{4,0}\rangle^{-}=|\downarrow\xi^{(i)}_{4,0}\rangle^{+}-|\uparrow\xi^{(i)}_{4,0}\rangle^{+}$,\vspace{.06in}\\
    $|\downarrow\psi^{(i)}_{4,0}\rangle^{-}=|0\rangle_{A_i}(|21\rangle-|32\rangle)_{A_{\overline{i+1}}A_{\overline{{i+2}}}}$,\\
    $|\uparrow\psi^{(i)}_{4,0}\rangle^{-}=|0\rangle_{A_i}(|31\rangle-|22\rangle)_{A_{\overline{i+1}}A_{\overline{{i+2}}}}$,\\
    $|\updownarrow\psi^{(i)}_{4,0}\rangle^{-}=|\downarrow\psi^{(i)}_{4,0}\rangle^{+}-|\uparrow\psi^{(i)}_{4,0}\rangle^{+}$,\vspace{.06in}\\
    $|\downarrow\eta^{(i)}_{4,0}\rangle^{-}=|0\rangle_{A_i}(|23\rangle-|34\rangle)_{A_{\overline{i+1}}A_{\overline{{i+2}}}}$,\\
    $|\uparrow\eta^{(i)}_{4,0}\rangle^{-}=|0\rangle_{A_i}(|33\rangle-|24\rangle)_{A_{\overline{i+1}}A_{\overline{{i+2}}}}$,\\
     $|\updownarrow\eta^{(i)}_{4,0}\rangle^{-}=|\downarrow\eta^{(i)}_{4,0}\rangle^{+}-|\uparrow\eta^{(i)}_{4,0}\rangle^{+}$,\vspace{.06in}\\
    $|\downarrow\kappa^{(i)}_{4,0}\rangle^{-}=|\updownarrow\phi^{(i)}_{4,0}\rangle^{+}-|\updownarrow\eta^{(i)}_{4,0}\rangle^{+}$,\\
    $|\uparrow\kappa^{(i)}_{4,0}\rangle^{-}=|\updownarrow\psi^{(i)}_{4,0}\rangle^{+}-|\updownarrow\xi^{(i)}_{4,0}\rangle^{+}$,\\
    $|\updownarrow\kappa^{(i)}_{4,0}\rangle^{-}=|\downarrow\kappa^{(i)}_{4,0}\rangle^{+}-|\uparrow\kappa^{(i)}_{4,0}\rangle^{+}$,\vspace{.15in}\\
    $|\kappa^{(i)}_{0,4}\rangle^{-}=|\updownarrow\kappa^{(\overline{i+1})}_{0,4}\rangle^{+}-|\updownarrow\kappa^{(\overline{i+2})}_{4,0}\rangle^{+}$,\\
    $|\kappa^{(i)}_{0,4}\rangle^{+}=|\updownarrow\kappa^{(\overline{i+1})}_{0,4}\rangle^{+}+|\updownarrow\kappa^{(\overline{i+2})}_{4,0}\rangle^{+}$,\\

    	\begin{figure}
    \begin{subfigure}{0.5\textwidth}
    	\includegraphics[width=3.3in,height=2.0in]{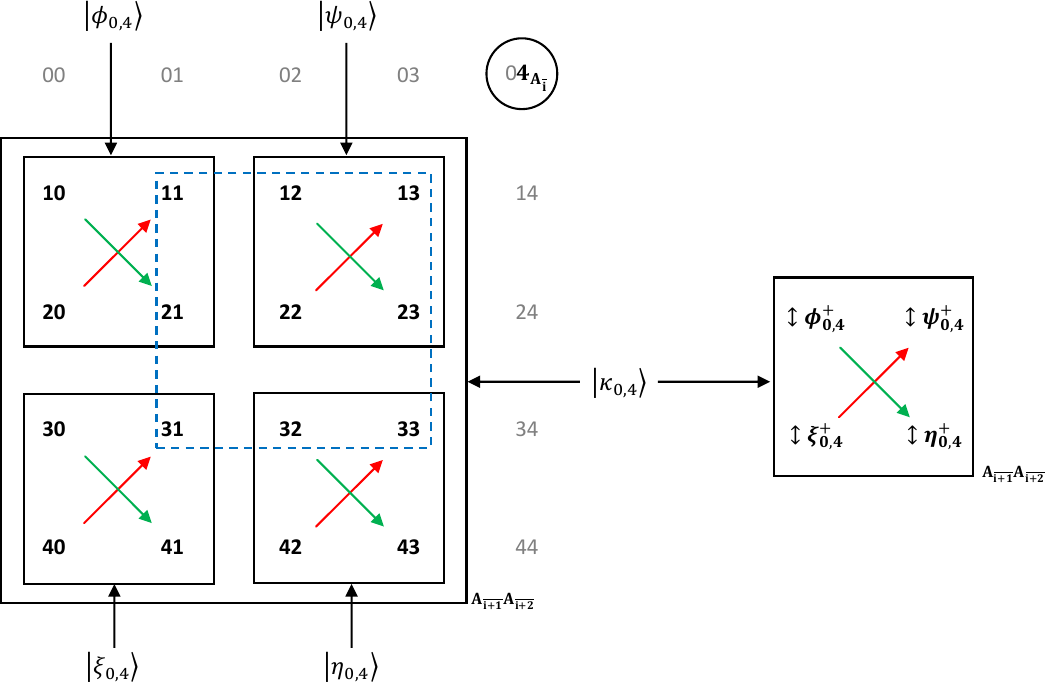}
    	\caption{}
    	\label{x12}
    \end{subfigure}
    \begin{subfigure}{0.5\textwidth}
    	\includegraphics[width=3.3in,height=2.0in]{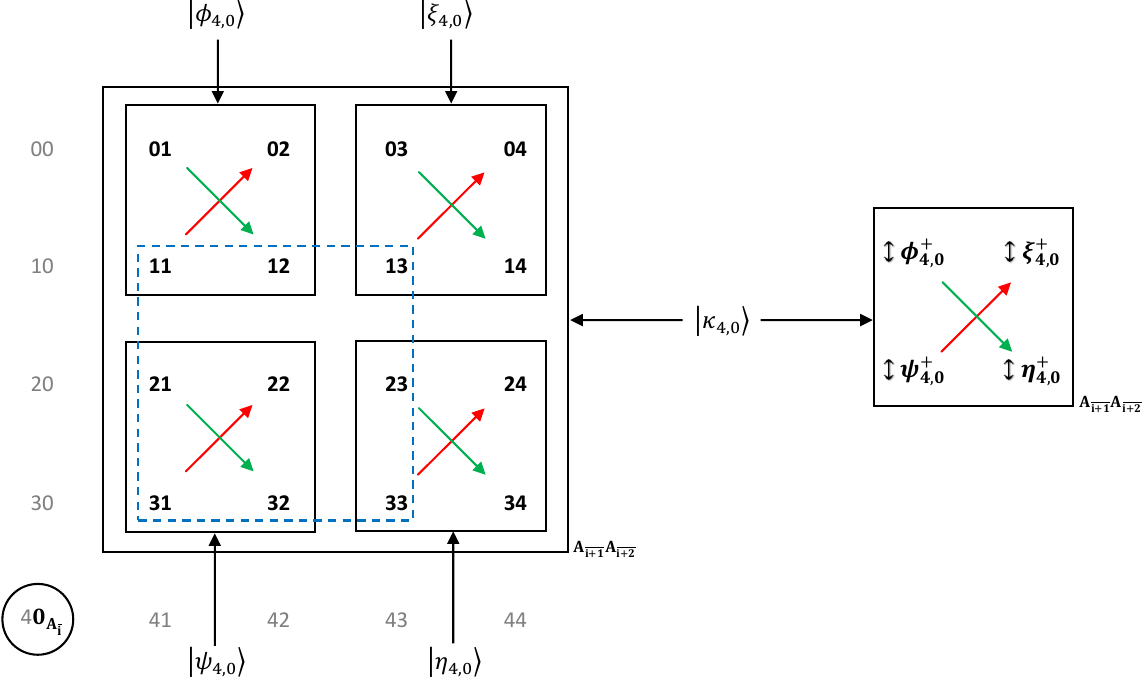}
    	\caption{}
    	\label{x13}
    \end{subfigure}
\begin{subfigure}{0.5\textwidth}
	\includegraphics[width=2.9in,height=1.3in]{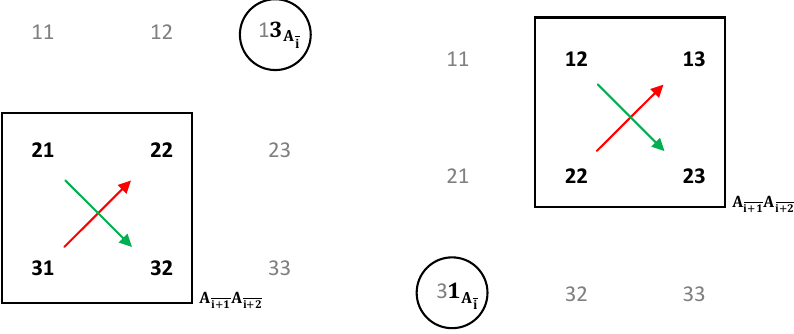}
	\caption{}
	\label{x13}
\end{subfigure}
    	\caption{Color outline for the construction of $\mathcal{UBB}_{II}$ in $\mathbb{C}^5\otimes\mathbb{C}^5\otimes\mathbb{C}^5$ as described above. The method is like peeling an onion. The upper two dominoes (sub-figures (a) and (b)) represent the outer part  whereas  the lower two (sub-figure (c)) represent the inner part of the Rubik cube representation for $\mathbb{C}^5\otimes\mathbb{C}^5\otimes\mathbb{C}^5$. For each element of the corresponding dominoes we can get a different $\mathcal{UBB}_{II}$ in $\mathbb{C}^5\otimes\mathbb{C}^5\otimes\mathbb{C}^5$. Here we choose the elements $40$ and $04$ for the outer part whereas $13$ and $31$ are chosen for the inner part.}
    	\label{2}
    \end{figure}

	The above 98 states along with the basis $\mathcal{B}^{1,3}_3$(considering \{$|1\rangle$,$|2\rangle$,$|3\rangle$\} be the ordered basis of Alice, Bob and Charlie) form a complete basis in $\mathbb{C}^5\otimes \mathbb{C}^5\otimes \mathbb{C}^5$ and we denote it as $\mathcal{B}^{0,4}_5$. Therefore every conjugate pair defines two complete orthogonal bases for the corresponding composite Hilbert space. Except for the diagonal elements(self conjugate) a $5 \otimes 5$ matrix contains 10 distinct pairs of conjugate elements. Then by this rule we can define 20 $\times$ 6(for $\mathcal{B}^{1,3}_3$)=120 orthogonal complete basis in $\mathbb{C}^{5}\otimes\mathbb{C}^{5}\otimes\mathbb{C}^{5}$.\\
	$|S\rangle=(|0\rangle+|1\rangle+|2\rangle+|3\rangle+|4\rangle)_A(|0\rangle+|1\rangle+|2\rangle+|3\rangle+|4\rangle)_B(|0\rangle+|1\rangle+|2\rangle+|3\rangle+|4\rangle)_C$
	be a stopper state. We claim that the set
	\begin{multline}
		$$
		\mathcal{U}^{0,4}_5=\mathcal{B}^{0,4}_5\cup\mathcal{B}^{1,3}_3 \cup\{|S\rangle\} \backslash\Bigl\{\left\{|\kappa^{(0)}_{0,4}\rangle^{+},|\kappa^{(1)}_{0,4}\rangle^{+},|\kappa^{(2)}_{0,4}\rangle^{+}\right\}\cup\\\left\{|\kappa^{(0)}_{1,3}\rangle^{+},|\kappa^{(1)}_{1,3}\rangle^{+},|\kappa^{(2)}_{1,3}\rangle^{+}\right\}
		\cup\left\{|\kappa_k\rangle\right\}_{k=0}^{4}\Bigl\}.
		$$
		\label{6}
	\end{multline}

	is a $\mathcal{UBB}_\mathcal{II}$ in $\mathbb{C}^{5}\otimes\mathbb{C}^{5}\otimes\mathbb{C}^{5}$.  The color outline for the method of construction is given in FIG.\ref{2}. Generalization of the construction for arbitrary local dimensions is presented in the next.$\blacksquare$\\
	
	 A distillable subspace must be with a negative partial transpose (NPT) subspace. In \cite{johnston2013} authors showed that for a $\mathbb{C}^{n_1} \otimes C^{n_2}$ system, dimension of distillable subspaces is upper bounded by $\left(n_1-\right.$ 1) $\left(n_2-1\right)$. Since any rank-4 bipartite NPT states are distillable \cite{linchen2016}, therefore when the composite system dimension is not more than 9, the NPT subspace is indeed a distillable subspaces and the explicit construction follows from Ref. [18]. Even if the constructions of NPT subspaces is known for arbitrary large dimensional systems \cite{johnston2013}, but the distillability of those subspaces remains unclear. In fact, in Ref. \cite{linchen2016} the authors have conjectured a bound NPT states of rank- 5. With the further continuation of the works mentioned above, the authors in \cite{agrwal2019} constructed a 5-dimensional subspace in tripartite system which is distillable across every bipartition. The next section will provide a detailed discussion of these specific types of subspaces.
	 \section{$\text {CONSTRUCTIONS IN }\left(\mathbb{C}^4\right)^{\otimes 3}$}
	 So in $\mathbb{C}^{4}\otimes\mathbb{C}^{4}\otimes\mathbb{C}^{4}$ we define a complete orthogonal basis as follows, denote it by $\mathcal{B}^{0,3}_4$.\\
	 (For $i=0,1,2$ and $w$ being the cube root of unity)
	 $|\downarrow\phi^{(i)}_{0,3}\rangle^{w}=|3\rangle_{A_i}(|10\rangle+{\omega}|21\rangle+{\omega}^2|32\rangle)_{A_{\overline{i+1}}A_{\overline{{i+2}}}}$,\\
	 $|\downarrow\phi^{(i)}_{0,3}\rangle^{w^2}=|3\rangle_{A_i}(|10\rangle+{\omega}^2|21\rangle+{\omega}|32\rangle)_{A_{\overline{i+1}}A_{\overline{{i+2}}}}$,\vspace{.06in}\\
	 $|\;\rotatebox[origin=c]{90}{$\hookrightarrow$}\;\phi^{(i)}_{0,3}\rangle^{w}=|3\rangle_{A_i}(|20\rangle+{\omega}|31\rangle+{\omega}^2|12\rangle)_{A_{\overline{i+1}}A_{\overline{{i+2}}}}$,\\
	 $|\;\rotatebox[origin=c]{90}{$\hookrightarrow$}\;\phi^{(i)}_{0,3}\rangle^{w^2}=|3\rangle_{A_i}(|20\rangle+{\omega}^2|31\rangle+{\omega}|12\rangle)_{A_{\overline{i+1}}A_{\overline{{i+2}}}}$,\vspace{.06in}\\
	  $|\Rsh\phi^{(i)}_{0,3}\rangle^{w}=|3\rangle_{A_i}(|30\rangle+{\omega}|11\rangle+{\omega}^2|22\rangle)_{A_{\overline{i+1}}A_{\overline{{i+2}}}}$,\\
	  $|\Rsh\phi^{(i)}_{0,3}\rangle^{w^2}=|3\rangle_{A_i}(|30\rangle+{\omega}^2|11\rangle+{\omega}|22\rangle)_{A_{\overline{i+1}}A_{\overline{{i+2}}}}$,\vspace{.1in}\\
	 $|\updownarrow\phi^{(i)}_{0,3}\rangle^{w}=|\downarrow\phi^{(i)}_{0,3}\rangle^{+}+\omega|\;\rotatebox[origin=c]{90}{$\hookrightarrow$}\;\phi^{(i)}_{0,3}\rangle^{+}+{\omega}^2|\Rsh\phi^{(i)}_{0,3}\rangle^{+}$,\\
	 $|\updownarrow\phi^{(i)}_{0,3}\rangle^{w^2}=|\downarrow\phi^{(i)}_{0,3}\rangle^{+}+{\omega}^2|\;\rotatebox[origin=c]{90}{$\hookrightarrow$}\;\phi^{(i)}_{0,3}\rangle^{+}+\omega|\Rsh\phi^{(i)}_{0,3}\rangle^{+}$,\\where,\\
	 \hspace*{.2in}$|\downarrow\phi^{(i)}_{0,3}\rangle^{+}=|3\rangle_{A_i}(|10\rangle+|21\rangle+|32\rangle)_{A_{\overline{i+1}}A_{\overline{{i+2}}}}$,\\
	 \hspace*{.2in}$|\;\rotatebox[origin=c]{90}{$\hookrightarrow$}\;\phi^{(i)}_{0,3}\rangle^{+}=|3\rangle_{A_i}(|20\rangle+|31\rangle+|12\rangle)_{A_{\overline{i+1}}A_{\overline{{i+2}}}}$,\\
	\hspace*{.2in}$|\Rsh\phi^{(i)}_{0,3}\rangle^{+}=|3\rangle_{A_i}(|30\rangle+|11\rangle+|22\rangle)_{A_{\overline{i+1}}A_{\overline{{i+2}}}}$,\vspace{.2in}\\
	 $|\downarrow\phi^{(i)}_{3,0}\rangle^{w}=|0\rangle_{A_i}(|01\rangle+{\omega}|12\rangle+{\omega}^2|23\rangle)_{A_{\overline{i+1}}A_{\overline{{i+2}}}}$,\\
	 $|\downarrow\phi^{(i)}_{3,0}\rangle^{w^2}=|0\rangle_{A_i}(|01\rangle+{\omega}^2|12\rangle+{\omega}|23\rangle)_{A_{\overline{i+1}}A_{\overline{{i+2}}}}$,\vspace{.06in}\\
	 $|\;\rotatebox[origin=c]{90}{$\hookrightarrow$}\;\phi^{(i)}_{3,0}\rangle^{w}=|0\rangle_{A_i}(|02\rangle+{\omega}|13\rangle+{\omega}^2|21\rangle)_{A_{\overline{i+1}}A_{\overline{{i+2}}}}$,\\
	 $|\;\rotatebox[origin=c]{90}{$\hookrightarrow$}\;\phi^{(i)}_{3,0}\rangle^{w^2}=|0\rangle_{A_i}(|02\rangle+{\omega}^2|13\rangle+{\omega}|21\rangle)_{A_{\overline{i+1}}A_{\overline{{i+2}}}}$,\vspace{.06in}\\
	 $|\Rsh\phi^{(i)}_{3,0}\rangle^{w}=|0\rangle_{A_i}(|03\rangle+{\omega}|11\rangle+{\omega}^2|22\rangle)_{A_{\overline{i+1}}A_{\overline{{i+2}}}}$,\\
	 $|\Rsh\phi^{(i)}_{3,0}\rangle^{w^2}=|0\rangle_{A_i}(|03\rangle+{\omega}^2|11\rangle+{\omega}|22\rangle)_{A_{\overline{i+1}}A_{\overline{{i+2}}}}$,\vspace{.1in}\\
	 $|\updownarrow\phi^{(i)}_{3,0}\rangle^{w}=|\downarrow\phi^{(i)}_{3,0}\rangle^{+}+\omega|\;\rotatebox[origin=c]{90}{$\hookrightarrow$}\;\phi^{(i)}_{3,0}\rangle^{+}+{\omega}^2|\Rsh\phi^{(i)}_{3,0}\rangle^{+}$,\\
	 $|\updownarrow\phi^{(i)}_{3,0}\rangle^{w^2}=|\downarrow\phi^{(i)}_{3,0}\rangle^{+}+{\omega}^2|\;\rotatebox[origin=c]{90}{$\hookrightarrow$}\;\phi^{(i)}_{3,0}\rangle^{+}+\omega|\Rsh\phi^{(i)}_{3,0}\rangle^{+}$,\\ where,\\
	 \hspace*{.2in}$|\downarrow\phi^{(i)}_{3,0}\rangle^{+}=|0\rangle_{A_i}(|01\rangle+|12\rangle+|23\rangle)_{A_{\overline{i+1}}A_{\overline{{i+2}}}}$,\\
	 \hspace*{.2in}$|\;\rotatebox[origin=c]{90}{$\hookrightarrow$}\;\phi^{(i)}_{3,0}\rangle^{+}=|0\rangle_{A_i}(|02\rangle+|13\rangle+|21\rangle)_{A_{\overline{i+1}}A_{\overline{{i+2}}}}$,\\
	 \hspace*{.2in}$|\Rsh\phi^{(i)}_{3,0}\rangle^{+}=|0\rangle_{A_i}(|03\rangle+|11\rangle+|22\rangle)_{A_{\overline{i+1}}A_{\overline{{i+2}}}}$,\vspace{.2in}\\
	 $|\phi^{(i)}_{0,3}\rangle^{\pm}=|\updownarrow\phi^{(i)}_{0,3}\rangle^{+}\pm|\updownarrow\phi^{(i)}_{3,0}\rangle^{+}$,\\ where,\\
	 \hspace*{.2in}$|\updownarrow\phi^{(i)}_{0,3}\rangle^{+}=|\downarrow\phi^{(i)}_{0,3}\rangle^{+}+|\;\rotatebox[origin=c]{90}{$\hookrightarrow$}\;\phi^{(i)}_{0,3}\rangle^{+}+|\Rsh\phi^{(i)}_{0,3}\rangle^{+}$,\\
	 \hspace*{.2in}$|\updownarrow\phi^{(i)}_{3,0}\rangle^{+}=|\downarrow\phi^{(i)}_{3,0}\rangle^{+}+|\;\rotatebox[origin=c]{90}{$\hookrightarrow$}\;\phi^{(i)}_{3,0}\rangle^{+}+|\Rsh\phi^{(i)}_{3,0}\rangle^{+}$,\vspace{.1in}\\
	 $|\psi_{0}\rangle=|1-2\rangle_{A_0}|1-2\rangle_{A_1}|1-2\rangle_{A_2}$,\\
	 $|\psi_{1}\rangle=|1+2\rangle_{A_0}|1-2\rangle_{A_1}|1-2\rangle_{A_2}$,\\
	 $|\psi_{2}\rangle=|1-2\rangle_{A_0}|1+2\rangle_{A_1}|1-2\rangle_{A_2}$,\\
	 $|\psi_{3}\rangle=|1-2\rangle_{A_0}|1-2\rangle_{A_1}|1+2\rangle_{A_2}$,\\
	 $|\psi_{4}\rangle=|1+2\rangle_{A_0}|1+2\rangle_{A_1}|1-2\rangle_{A_2}$,\\
	 $|\psi_{5}\rangle=|1+2\rangle_{A_0}|1-2\rangle_{A_1}|1+2\rangle_{A_2}$,\\
	 $|\psi_{6}\rangle=|1-2\rangle_{A_0}|1+2\rangle_{A_1}|1+2\rangle_{A_2}$,\\
	 $|\psi_{7}\rangle=|1+2\rangle_{A_0}|1+2\rangle_{A_1}|1+2\rangle_{A_2}$,\vspace{.1in}\\
	 $\left\{|\kappa_{k}\rangle=|k\rangle_{A_0}|k\rangle_{A_1}|k\rangle_{A_2},k=0,3\right\}$\\
	 
\begin{figure}
 	\centering
 	\includegraphics[width=.46\textwidth]{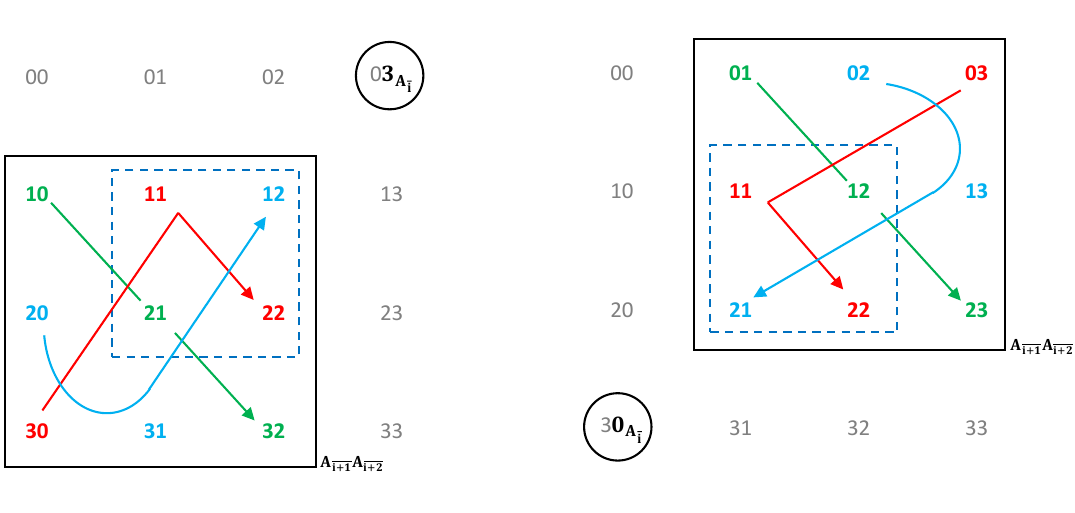}
 	\caption{Color outline for the construction of $\mathcal{UBB}_{II}$ in $\mathbb{C}^4\otimes\mathbb{C}^4\otimes\mathbb{C}^4$ as described above. The green down-arrow relates with $|\downarrow\phi^{(i)}_{0,3}\rangle$, whereas the sky hook-arrow and the red shift-arrow relates with $|\;\rotatebox[origin=c]{90}{$\hookrightarrow$}\;\phi^{(i)}_{0,3}\rangle$ and $|\Rsh\phi^{(i)}_{0,3}\rangle$ respectively.}
 	\label{3}
 \end{figure}

	 The 64 states described above form a complete basis of $\mathbb{C}^4\otimes\mathbb{C}^4\otimes\mathbb{C}^4$. Let,
	 $|S\rangle=(|0\rangle+|1\rangle+|2\rangle+|2\rangle)_A(|0\rangle+|1\rangle+|2\rangle+|2\rangle)_B(|0\rangle+|1\rangle+|2\rangle+|2\rangle)_C$
	 be a stopper state. We claim that the set
	 \begin{multline}
	 	$$
	 	\mathcal{U}^{0,3}_4=\mathcal{B}^{0,3}_4 \cup\{|S\rangle\} \backslash\Bigl\{\left\{|\phi^{(0)}_{0,3}\rangle^{+},|\phi^{(1)}_{0,3}\rangle^{+},|\phi^{(2)}_{0,3}\rangle^{+},|\psi_7\rangle\right\}
	 	\\\cup\left\{|\kappa_k\rangle\right\}_{k=0,3}\Bigl\}.
	 	$$
	 	\label{51}
	 \end{multline}
 	is a $\mathcal{UBB}_\mathcal{II}$ in $\mathbb{C}^{4}\otimes\mathbb{C}^{4}\otimes\mathbb{C}^{4}$.  The color outline for the method of construction is given in FIG.\ref{3}. Generalization of the construction for arbitrary local dimensions is presented in the next section.$\blacksquare$\\
 \vspace{.1in}
	 \section{Construction of $\mathcal{UBB}_{II}$ in arbitrary large dimension}
\begin{figure}
	 	\begin{subfigure}{0.5\textwidth}
	 		\includegraphics[width=3.5in,height=1.55in]{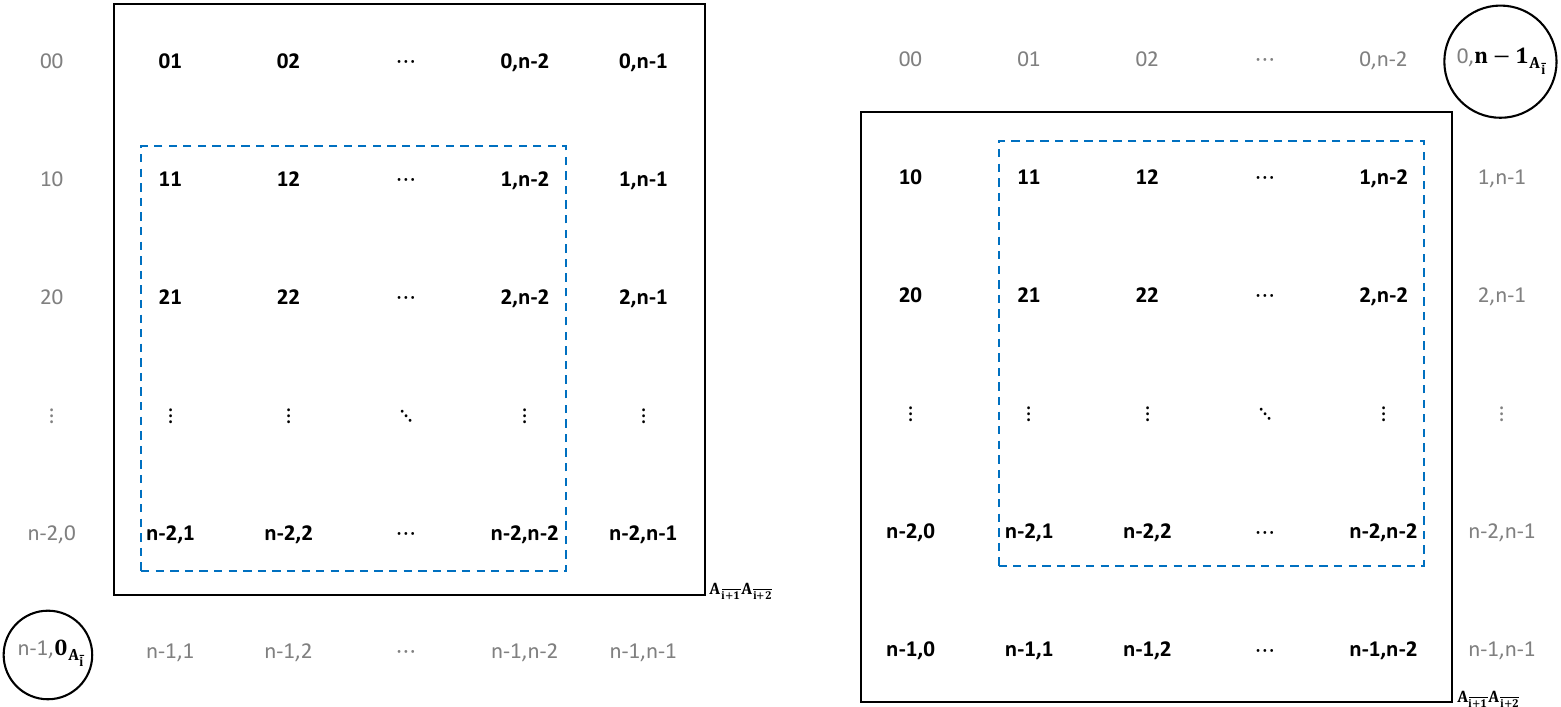}
	 		\caption{}
	 		\label{x12}
	 	\end{subfigure}
	 	\begin{subfigure}{0.5\textwidth}
	 		\includegraphics[width=3.5in,height=1.55in]{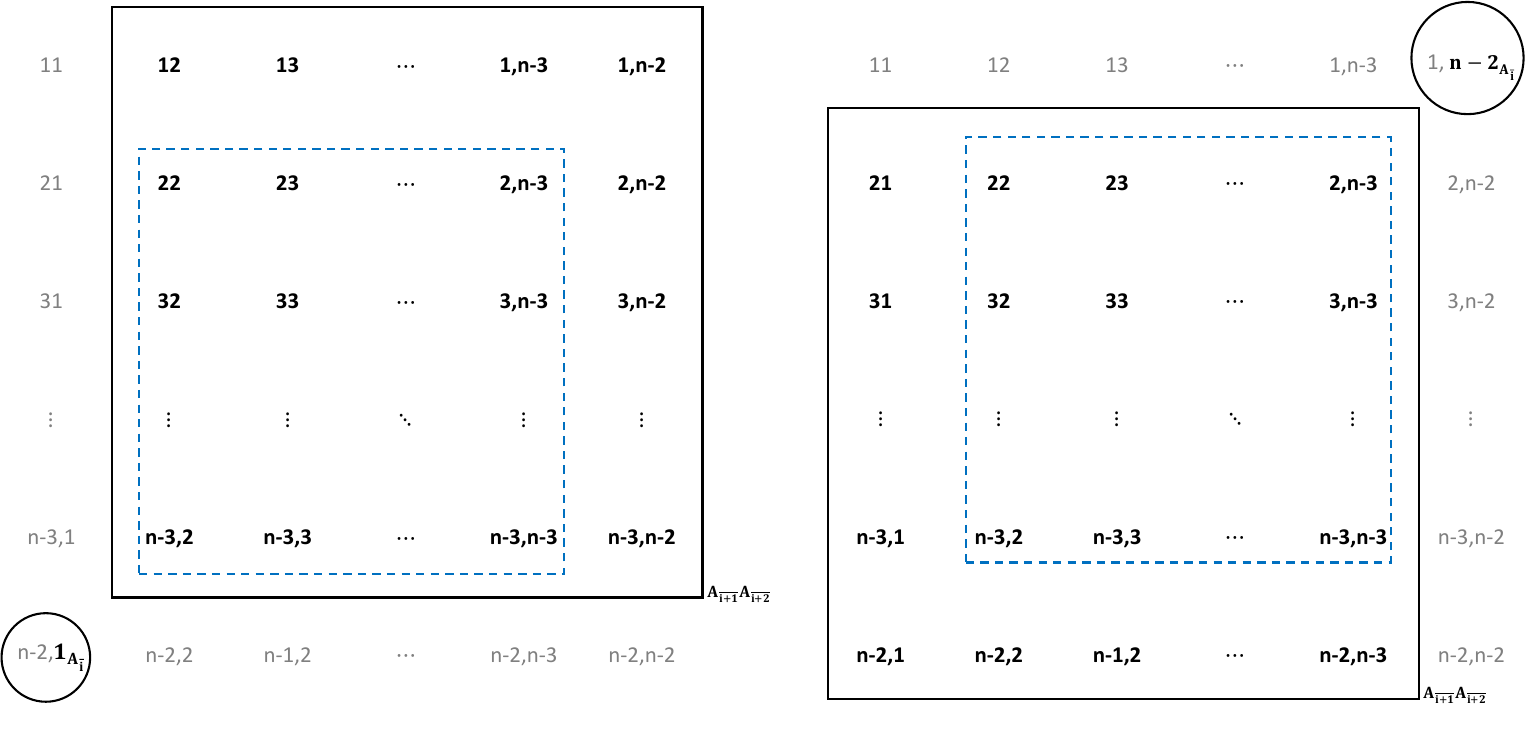}
	 		\caption{}
	 		\label{x13}
	 	\end{subfigure}
	 	\begin{subfigure}{0.5\textwidth}
	 		\includegraphics[width=3.5in,height=1.5in]{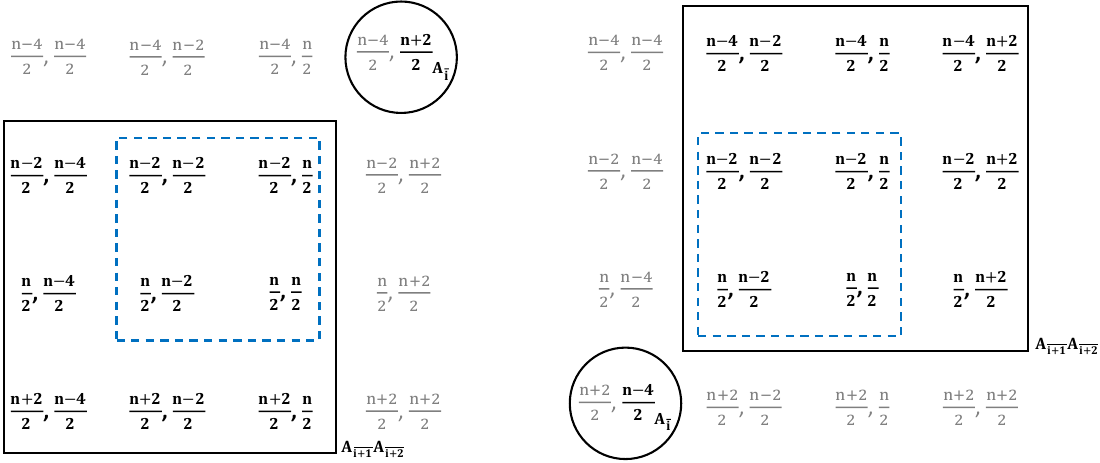}
	 		\caption{}
	 		\label{x13}
	 	\end{subfigure}
 	\begin{subfigure}{0.5\textwidth}
 		\includegraphics[width=3.5in,height=1.5in]{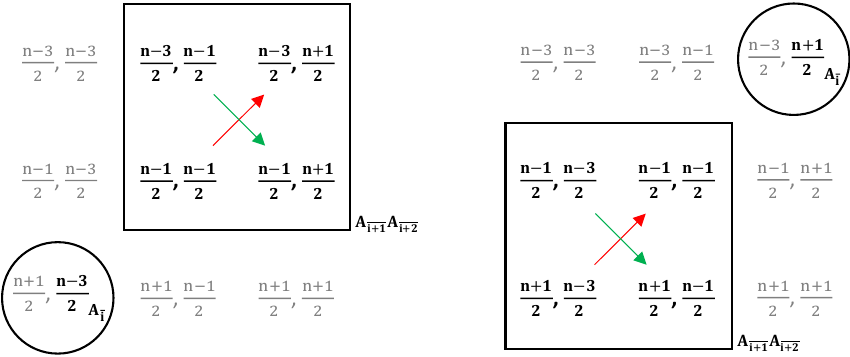}
 		\caption{}
 		\label{x13}
 	\end{subfigure}
	 	\caption{Color outline for the construction of $\mathcal{UBB}_{II}$ in $\mathbb{C}^n\otimes\mathbb{C}^n\otimes\mathbb{C}^n$ as described above. The method is like peeling an onion. Sub-figure (a) defines for the outer most layer. Sub-figure (b) and Sub-figure (c) next inner layer respectively. By continuing this Sub-figure (d) represents for the inner most layer. For each element of the corresponding dominoes we can get a different $\mathcal{UBB}_{II}$ in $\mathbb{C}^5\otimes\mathbb{C}^5\otimes\mathbb{C}^5$. Here we choose the elements $(n-1)0$ and $0(n-1)$ for the outer part.}
	 	\label{4}
	 \end{figure}

	 \begin{widetext}
	 	In this section our aim is to generalize the whole structure discussed earlier in $n\otimes n\otimes n, n\geq3$. The stopper state for this dimension is
	 	\begin{equation}
	 		\bigotimes\limits_{i=1}^{3}\left[\sum\limits_{k=0}^{n-1}\ket{k}\right]_{A_{\overline{i}}}
	 	\end{equation}
	 	
	 	We will choose the rest bi-separable states from the outer part of the cube. Then we will go for the inner part which is nothing but another $n-2\otimes n-2\otimes n-2$ cube and we will choose states from the outer part of the recent cube and so on. And the inner most cube is of  $3\otimes 3\otimes 3$ (odd $n$) and $4\otimes 4\otimes 4$ (even $n$). To define it mathematically we choose a variable say $s_n=0(1)(\lfloor \frac{n}{2} \rfloor-1)$. $s_n=0$ corresponds to the outer most part where as $s_n=\lfloor \frac{n}{2} \rfloor-1$ corresponds to the innermost part of the $n\otimes n\otimes n$ cube.
	 	
	 	Now we fixed $s_n=s$.The states are
	 	\begin{equation}
	 		\left[\sum\limits_{k=s}^{(n-2)-s}\ket{k}\right]_{A_{\overline{i}}}\otimes\left[\ket{(n-1)-s}\otimes\left\{\sum\limits_{k=s+1}^{(n-2)-s}\ket{k}\right\}-\left\{\sum\limits_{k=s+1}^{(n-2)-s}\ket{k}\right\}\otimes\ket{s}\right]_{A_{\overline{i+1}}A_{\overline{i+2}}}
	 	\end{equation}

	 	Now $(n-1)-2s_n$ can be factorized in $r$ prime factors $p_1,p_2,\cdots,p_r$ with $2\leq p_1 \leq p_1 \leq \cdots \leq p_r$ i.e., $(n-1)-2s=p_1p_2 \cdots p_r$. We now define $\rho_l=\prod\limits_{i=0}^{l}p_i$ and $\alpha_l=\frac{(n-1)-2s}{\rho(l)}$. Then we introduce another variable $l_s=0(1)(r-1)$.

	 	Fix $l_s=l$. Now introduce new four variables $d_l=0(1)(\alpha_{l+1}-1)$,$h_l=0(1)(\alpha_{l+1}-1)$, $j_l=1(1)(p_{l+1}-1)$ and $t_l=0(1)(p_{l+1}-1)$.
	 	
	 	For $d_l=d$,$h_l=h$, $j_l=j$ and $t_l=t$ we define
	 	
	 	\begin{multline*}
	 		\ket{\kappa_{d,h,j,t}^{n,s,l}}_{BC}=\sum\limits_{k=0}^{p_{l+1}-t-1}w_{j,k}^{p_{l+1}}\left \{ \sum \limits_{m=0}^{\rho_l-1}\ket{(s+1)+d\rho_{l+1}+k\rho_l+m}  \right \}_{B}\otimes \left \{ \sum \limits_{m=0}^{\rho_l-1}\ket{s+h\rho_{l+1}+(k+t)\rho_l+m}  \right \}_{C}
	 		\\ 
	 		+\sum\limits_{k=0}^{t-1}w_{j,p_{l+1}-t+k}^{p_{l+1}}\left \{ \sum \limits_{m=0}^{\rho_l-1}\ket{(s+1)+(d+1)\rho_{l+1}+(k-t)\rho_l+m}  \right \}_{B}\otimes\left \{ \sum \limits_{m=0}^{\rho_l-1}\ket{s+h\rho_{l+1}+k\rho_l+m}  \right \}_{C}
	 	\end{multline*}

	 	The states are
	 	\begin{equation}
	 		\ket{(n-1)-s}_{A_{\overline{i}}}\otimes\ket{\kappa_{d,h,j,t}^{n,s,l}}_{A_{\overline{i+1}}A_{\overline{i+2}}}
	 	\end{equation}

	 	and
	 	\begin{equation}
	 		\ket{s}_{A_{\overline{i}}}\otimes\ket{\kappa_{d,h,j,t}^{n,s,l}}_{A_{\overline{i+2}}A_{\overline{i+1}}}
	 	\end{equation}
	 	
	 	For $d_l=d$,$h_l=h$ and $j_l=j$, we define
	 	$$
	 	\ket{\kappa_{d,h,j}^{n,s,l}}_{BC}=\sum\limits_{t_l=0}^{p_{l+1}-1}w_{j,t_l}^{p_{l+1}}\ket{\kappa_{d,h,0,t_l}^{n,s,l}}_{BC}
	 	$$
	 	
	 	The states are
	 	\begin{equation}
	 		\ket{(n-1)-s}_{A_{\overline{i}}}\otimes\ket{\kappa_{d,h,j}^{n,s,l}}_{A_{\overline{i+1}}A_{\overline{i+2}}}
	 	\end{equation}
	 	
	 	and
	 	\begin{equation}
	 		\ket{s}_{A_{\overline{i}}}\otimes\ket{\kappa_{d,h,j}^{n,s,l}}_{A_{\overline{i+2}}A_{\overline{i+1}}}
	 	\end{equation}

	 	$$
	 	w_{j,k}^{p}=e^{jk\frac{2\pi}{p}i},i=\sqrt{-1}.
	 	$$
	 	
	 \end{widetext}
	 Here we generalize the $\mathcal{UBB}_\mathcal{II}$ for arbitrary large composite Hilbert space $\mathbb{C}^n\otimes\mathbb{C}^n\otimes\mathbb{C}^n$. Using the same method of construction as described above we can get a chain of matrix pair with descending dimensions. The color outline for the method of construction is given in FIG.\ref{4}.
\section{construction of genuinely entangled space and possible application}
	The complementary subspace of $\mathcal{U}^{0,2}_3$ contains no biseparable states, thus it becomes a genuinely entangled space(GES). In \cite{agrwal2019} author constructed a GES which is distillable across every bipartition. So it is quite interesting to check whether the complementary subspace of $\mathcal{U}^{0,2}_3$ is bidistillable or not. Suppose $\mathcal{P}(n)$ be the rank $n$ projector $(1\leq n \leq 5)$ acting on $\mathbb{GE}(5)$ and any state of $\mathbb{GE}(5)$ must be expressed as a linear combination of states from the set $\mathcal{K}=\Bigl\{\{|\kappa^{(0)}_{0,2}\rangle^{+},|\kappa^{(1)}_{0,2}\rangle^{+},|\kappa^{(2)}_{0,2}\rangle^{+}\}
	\cup\{|\kappa_k\rangle\}_{k=0}^{2}\}\Bigl\}$. Now to construct $n$ mutually orthogonal vectors in $\mathbb{GE}(5)$, we need at least $(n+1)$ states from $\mathcal{K}$ and the bimarginals obtained from the projector of those $n$ states must be of rank $(n+1)$. This proves the distillability of the states proportional to $\mathbb{P}(n)$ across every bipartition \cite{agrwal2019}. It follows from the fact that for a $n$-dimensional subspace $S_{\alpha\beta}$ of $\mathbb{C}^{d_\alpha}\bigotimes\mathbb{C}^{d_\beta}$, if the projector $\mathbb{P}_{\alpha\beta}$ on $S_{\alpha\beta}$ satisfies $\mathcal{R}(\mathbb{P}_{\alpha\beta})<max\Bigl\{\mathcal{R}(\mathbb{P}_{\alpha}), \mathcal{R}(\mathbb{P}_{\beta})\Bigl\}$, then all rank-$n$ states supported on $S_{\alpha\beta}$ are $1-$copy distillable where $\mathbb{P}_{\alpha(\beta)}=Tr_{\beta(\alpha)}(\mathbb{P}_{\alpha\beta})$ and $\mathcal{R}(.)$ denotes the rank of the operator.
\vspace{.5cm}\\
The state $\rho_{\mathbb{GE}}^{S} (5)$ proportional to the projector on the subspace $\mathbb{GE}(5)$ is given by \\
\begin{center}
	$\rho_{\mathbb{GE}}^{S} (5):= \frac{1}{5}\Bigl(\mathbb{I}_3 \otimes \mathbb{I}_3 \otimes\mathbb{I}_3 - \sum_{\chi\in\mathcal{U}^{0,2}_3}{|\bar{\chi}\rangle\langle\bar{\chi}|} \Bigl) $
\end{center}

Here, $\ket{\Bar{\chi}}$ is the normalized state proportional to $\ket{\chi}$. Since the construction is party symmetric, all the two party reduced states $\rho_{\beta}:= Tr_{\alpha}[\rho_{\mathbb{GE}}^{S}(5)]$, with $\beta\in \{ BC, CA, AB \} $ and $\alpha\in\{ A, B, C\}$ respectively, are identical and the corresponding density matrix takes the following form:
\begin{center}
	$$
	\rho_\beta = \frac{1}{360}
	\begin{pmatrix}
		82 & 10 & -8 & 1 & 1 & -8 & 1 & 1 & -8 \\
		10 & 19 & 1 & 1  & 10 & 1 & 1 & 1 & -8 \\
		-8 & 1 & 19 & -8  & 1 & 19 & -8 & -8 & -8 \\
		1 & 1 & -8 & 19 & 1 & -8 & 19 & 1 & -8\\
		1 & 10 & 1 & 1 & 82 & 1 & 19 & 1 & -8\\
		-8 & 1 & 19 & -8 & 1 & 19 & 10 & -8 & -8\\
		1 & 1 & -8 & 19 & 19 & 10 & 19 & 19 & 10\\
		1 & 1 & -8 & 1 & 1 & -8 & 19 & 19 & 10\\
		-8 & -8 & -8 & -8 & -8 & -8 & 10 & 10 & 82\\
		
	\end{pmatrix}
	$$
\end{center}
The rank of $\rho_\beta$ is 6. Therefore the state $\rho_{\mathbb{GE}}^{S} (5)$ is bidistillable across every bipartition.\\\\
	It is possible to construct strongly nonlocal  $\mathcal{UBB}_\mathcal{II}$ in $\mathbb{C}^{n_1} \otimes \mathbb{C}^{n_2} \otimes\mathbb{C}^{n_3}$ by  applying the same procedure. However, in that case we need to face several difficulties. First we need to construct $\mathcal{UBB}_\mathcal{II}$ in $\mathbb{C}^{n_1} \otimes \mathbb{C}^{n_2} \otimes\mathbb{C}^{n_3}$. Second for the strong quantum nonlocality of the $\mathcal{UBB}_\mathcal{II}$, we require that it has a similar structure under cyclic permutation of the subsystems. Otherwise, we need to show that any two subsystems can only perform a trivial orthogonality-preserving POVM, and it causes a lot of calculations.\\
	It is known that UPB is locally indistinguishable and so $\mathcal{UBB}_\mathcal{I}$. It is quite interesting to know whether there exists a $\mathcal{UBB}_\mathcal{II}$ which is locally indistinguishable. In [43] author constructed a strong nonlocal UPB which can be trivially extended to a strongly nonlocal $\mathcal{UBB}_\mathcal{I}$. One may ask whether there exists a $\mathcal{UBB}_\mathcal{II}$ which is strongly nonlocal. In this paper, we solve this problem in a different way.
	The application of the strong nonlocal $\mathcal{UBB}_\mathcal{II}$ in secret sharing could be as follows:
	
	 Suppose that some secret information encoded by a bunch of orthogonal quantum states(some of which contain bipartite entanglement also) is shared between three parties Alice, Bob, and Charlie. So for any pair of parties, the average correlation between them is not exactly equal to zero. The task is to decode the information together at some future stage. It is also restricted that any operation which gives the final failure for future decoding of information might not be allowed. So it is assumed that every participant only can perform orthogonality-preserving local measurement(OPLM). Otherwise, even the global measurement would not be able to decode the information in the future. The concept of strong quantum nonlocality guarantees the security of the encoded information like no one of Alice, Bob, and Charlie can reveal the information by OPLM even if any two of them come closure the security of the information remain same.\\  
    \section{discussion}
    \label{A5}
    Multipartite entanglement is a fundamental concept in quantum physics that describes the intricate correlations among multiple quantum particles. It has practical applications in quantum technologies and plays a vital role in understanding complex quantum systems\cite{Hillery,briegel2001,komer2014}.  An UBB is a set of orthogonal pure biseparable states which span a subspace of a given Hilbert space while the complementary subspace contains only genuinely entangled states. The importance of UBB lies in its ability to define a subspace of the Hilbert space that contains only genuinely entangled states. Genuine entanglement is crucial in quantum information processing because it signifies strong correlations that are not reducible to classical probabilistic models. Here we have established connections between the concept of unextendible biseparable bases and the phenomenon of strong quantum nonlocality in an extensive tripartite scenario. In fact, we are able to set up a wide class of UBB in $\mathbb{C}^n\otimes\mathbb{C}^n\otimes\mathbb{C}^n, d\geq3$ that does not contain any UPB as a proper subset of it. Specifically, we have shown that the above class of UBB satisfies the phenomenon of strong quantum nonlocality from the perspective of local elimination. The notion of UBB studied here is significant as it sufficiently leads to a subspace containing only genuinely entangled states. Our symmetric class $\mathcal{UBB}_{II}$ leads to a subspace that is not only a GES but also distillable across every bipartition. Our study also motivates some interesting questions for further research. First of all, the construction of a multipartite subspace is possible where the subspace is not only distillable across all bipartition but also distillable in partitions across multiple parties. Another important question is the relationship between strong quantum nonlocality and UBB in multipartite cases. From our knowledge, most of the references focus on the smallest number of states to show strong nonlocality, but generally speaking, we need enough states to show strong quantum nonlocality. So it is better to search for other methods to explain the relationship between strong quantum nonlocality and UBB in the future.	

	\section*{ACKNOWLEDGEMENTS}
	The authors I. Chattopadhyay and D. Sarkar acknowledge the work as part of QUest initiatives by DST India. The authors A. Bhunia and I. Biswas acknowledge the support from UGC, India. The author S. Bera acknowledges the support from CSIR, India. 
	
\section*{Appendix A: Proof of Theorem 1.}
We only need to show any party cannot start a non-trivial OPLM. As we see that the states in (\ref{5}) follow the cyclic property, then if any one party (say party $A_0$) goes first and cannot start a nontrivial measurement then parties $A_1$ and $A_2$ also cannot start nontrivial OPLM. So, it is sufficient to prove $A_0$ can only perform the measurement proportional to the identity.\\
Suppose $ E_{\alpha}^{(0)}=M_{\alpha}^{\dag}M_{\alpha} $'s are such measurements that $A_0$ start. As $A_0$'s system is defined in three dimensional Hilbert space $\mathcal{H}^{(0)}$ in $ \{\ket{0}, \ket{1}, \ket{2} \}_{A_0} $ basis, we can write $ E_{\alpha}$ as $3 \cross 3 $ square matrix, as follows:
\begin{equation}
	E_{\alpha}^{(0)}=
	\bordermatrix{
		~ & \bra{0} &\bra{1} & \bra{2} \cr
		\ket{0} & \alpha_{00} & \alpha_{01} & \alpha_{02} \cr
		\ket{1} &\alpha_{10} & \alpha_{11} & \alpha_{12} \cr
		\ket{2} &\alpha_{20} & \alpha_{21} & \alpha_{22}\cr}
\end{equation}
After measurement, all the states either eliminate or remains orthogonal. In both cases $ \mel{\phi}{E_{\alpha}^{(0)} \otimes I_{3}^{(1)} \otimes I_{3}^{(2)}}{\psi} = 0 , \phi \neq \psi ,\phi, \psi \in \mathcal{U}_3^{0,2} $ and for every outcome $ \alpha $.
Then considering the pairs  $ |\downarrow\kappa^{(0)}_{2,0}\rangle^{-}, |\uparrow\kappa^{(2)}_{0,2}\rangle^{-}$ we get 
$$
\begin{array}{l}
	^{-}\left\langle \downarrow\kappa^{(0)}_{2,0}\left|E_{\alpha}^{(0)} \otimes I_{3}^{(1)} \otimes I_{3}^{(2)}\right|\uparrow\kappa^{(2)}_{0,2}\right\rangle^{-} = 0 \text{,}\\
	^{-}\left\langle \uparrow\kappa^{(2)}_{0,2}\left|E_{\alpha}^{(0)} \otimes I_{3}^{(1)} \otimes I_{3}^{(2)}\right|\downarrow\kappa^{(0)}_{2,0}\right\rangle^{-} = 0
\end{array}
$$
i.e.,
$$
\begin{array}{l}
	\mel{0}{E_{\alpha}}{1}_{A_0} \braket{1}{1}_{A_1}\braket{2}{2}_{A_2} = 0  \text{,}\\
	\mel{1}{E_{\alpha}}{0}_{A_0} \braket{1}{1}_{A_1}\braket{2}{2}_{A_2} = 0 
\end{array}
$$
i.e.,
\begin{equation}
	\alpha_{01}= \alpha_{10}=0
	\label{ee1}
\end{equation}
Similarly, for pairs $|\uparrow\kappa^{(1)}_{0,2}\rangle^{-}, |\downarrow\kappa^{(0)}_{0,2}\rangle^{-}$ and $ |\downarrow\kappa^{(0)}_{2,0}\rangle^{-}, |\downarrow\kappa^{(1)}_{2,0}\rangle^{-}$ we get 
\begin{equation}
	\alpha_{12}= \alpha_{21}=0
	\label{ee2}
\end{equation}
\begin{equation}
	\alpha_{20}= \alpha_{02}=0
	\label{ee3}
\end{equation}
respectively.\\
Now the matrix $E_{\alpha}$ reduced to a diagonal matrix
\begin{equation}
	E_{\alpha}^{A}=
	\begin{pmatrix}
		\alpha_{00} & 0 & 0 \\
		0 & \alpha_{11} & 0 \\
		0 & 0 & \alpha_{22}
	\end{pmatrix}
\end{equation}
To show it is proportional to identity we only have to prove  $ \alpha_{00}= \alpha_{11}= \alpha_{22}$.\\
Now by choosing the pairs  $ \ket{S}, |\downarrow\kappa^{(1)}_{2,0}\rangle^{-}$ and $ \ket{S}, |\downarrow\kappa^{(2)}_{2,0}\rangle^{-}$, we get
$$
\begin{array}{l}
	\left\langle{S}\left|E_{\alpha}^{(0)} \otimes I_{3}^{(1)} \otimes I_{3}^{(2)}\right|\downarrow\kappa^{(1)}_{2,0}\right\rangle^{-}= 0 \text{,}\\
	\left\langle{S}\left|E_{\alpha}^{(0)} \otimes I_{3}^{(1)} \otimes I_{3}^{(2)}\right|\downarrow\kappa^{(2)}_{2,0}\right\rangle^{-}= 0
\end{array}
$$
implies,
\begin{multline}
	$$
	\mel{0+1+2}{E_{\alpha}}{1}_{A_0} \braket{0}{0}_{A_1}\braket{0}{0}_{A_2} -\\ \mel{0+1+2}{E_{\alpha}}{2}_{A_0} \braket{0}{0}_{A_1}\braket{1}{1}_{A_2}=0 \text{,}\\
	\mel{0+1+2}{E_{\alpha}}{0}_{A_0} \braket{1}{1}_{A_1}\braket{0}{0}_{A_2} -\\ \mel{0+1+2}{E_{\alpha}}{1}_{A_0} \braket{2}{2}_{A_1}\braket{0}{0}_{A_2}=0\\
	$$
\end{multline}
i.e.,
\begin{multline}
	$$
	\alpha_{01}+\alpha_{11}+\alpha_{21}= \alpha_{02}+\alpha_{12}+\alpha_{22},\\
	\alpha_{00}+\alpha_{10}+\alpha_{20}= \alpha_{01}+\alpha_{11}+\alpha_{21}
	$$
	\label{ee4}
\end{multline}
Now, from (\ref{ee1}),(\ref{ee2}),(\ref{ee3}) and (\ref{ee4}) we get 
\begin{equation}
	\alpha_{00}= \alpha_{11}=\alpha_{22}
\end{equation}
This completes the proof.$\blacksquare$
\section*{Appendix B: Proof of Theorem 2.}
 As the set of states given in (\ref{5}) is cyclic in every tripartition, it is also cyclic in every bipartition. So we only need to prove the states are irreducible in $A_0A_1|A_2$, i.e., parties $A_0$ and $A_1$ can apply joint measurement on the subsystem $A_0A_1$.\\
For that we rewrite the states in (\ref{5}) in the basis $ \{\ket{0}, \ket{1}, \ket{2}, \ket{3}, \ket{4}, \ket{5}, \ket{6}, \ket{7}, \ket{0} \}_{A_0A_1} $ instead of $ \{\ket{00}, \ket{01}, \ket{10}, \ket{20}, \ket{11}, \ket{02}, \ket{12}, \ket{21}, \ket{22} \}_{A_0A_1} $ respectively, as follows:
$$
\begin{array}{ll}
	|\downarrow \kappa^{(0)}_{0,2}\rangle^{-}= \ket{7}_{\overline{A_0A_1}}\ket{0}_{A_2}-\ket{8}_{\overline{A_0A_1}}\ket{1}_{A_2},\\
	|\downarrow \kappa^{(1)}_{0,2}\rangle^{-} = \ket{51}_{\overline{A_0A_1}A_2} - \ket{62}_{\overline{A_0A_1}A_2},\\
	|\downarrow \kappa^{(2)}_{0,2}\rangle^{-} = \ket{22} - \ket{72},\vspace{.06in}\\
	|\uparrow \kappa^{(0)}_{0,2}\rangle^{-} = \ket{80} - \ket{71},\\
	|\uparrow \kappa^{(1)}_{0,2}\rangle^{-} = \ket{52} - \ket{61},\\
	|\uparrow \kappa^{(2)}_{0,2}\rangle^{-} = \ket{32} - \ket{42},\vspace{.06in}\\
	|\downarrow \kappa^{(0)}_{2,0}\rangle^{-} = \ket{01} - \ket{12},\\
	|\downarrow \kappa^{(1)}_{2,0}\rangle^{-} = \ket{20} - \ket{31},\\
	|\downarrow \kappa^{(2)}_{2,0}\rangle^{-} = \ket{10} - \ket{60},\vspace{.06in}\\
	|\uparrow \kappa^{(0)}_{2,0}\rangle^{-} = \ket{11} - \ket{02},\\
	|\uparrow \kappa^{(1)}_{2,0}\rangle^{-} = \ket{21} - \ket{30},\\
	|\uparrow \kappa^{(2)}_{2,0}\rangle^{-} = \ket{40} - \ket{50},\vspace{.06in}\\
	|\updownarrow \kappa^{(0)}_{0,2}\rangle^{-} = \ket{7-8}\ket{0-1},\\
	|\updownarrow \kappa^{(1)}_{0,2}\rangle^{-} = \ket{5-6}\ket{1-2},\\
	|\updownarrow \kappa^{(2)}_{0,2}\rangle^{-}=\ket{2-3-4+7}\ket{2},\vspace{.06in}\\
	|\updownarrow \kappa^{(0)}_{2,0}\rangle^{-} = \ket{0-1}\ket{1-2},\\
	|\updownarrow \kappa^{(1)}_{2,0}\rangle^{-}=\ket{2-3}\ket{0-1},\\
	|\updownarrow \kappa^{(2)}_{2,0}\rangle^{-}=\ket{1-4-5+6}\ket{0},\vspace{.06in}\\
	|\kappa^{(0)}_{0,2}\rangle^{-} =\ket{5+6}\ket{1+2}-\ket{1+4+5+6}\ket{0},\\
	|\kappa^{(1)}_{0,2}\rangle^{-} = \ket{2+3+4+7}\ket{2}-\ket{0+1}\ket{1+2},\\
	|\kappa^{(2)}_{0,2}\rangle^{-} =\ket{7+8-2-3}\ket{0+1},\vspace{.06in}\\
	\ket{S} = \ket{0+1+2+3+4+5+6+7+8}\ket{0+1+2}.
\end{array}
$$
The proof is quite similar to the previous one. $A_0A_1$ starts an OPLM $ E_{\alpha}^{A_0A_1}=M_{\alpha}^{\dag}M_{\alpha} $ which is nothing but a square matrix of order 9.
\begin{equation}
	E_{\alpha}^{A_0A_1}=
	\bordermatrix{
		~ & \bra{0} &\bra{1} & \cdots & \bra{8} \cr
		\ket{0} & \alpha_{00} & \alpha_{01} & \cdots & \alpha_{08} \cr
		\ket{1} &\alpha_{10} & \alpha_{11} & \cdots & \alpha_{18} \cr
		\vdots & \vdots & \vdots & \ddots & \vdots \cr
		\ket{8} &\alpha_{80} & \alpha_{81} & \cdots & \alpha_{88}\cr}
\end{equation}
As we know $ M_{\alpha} \otimes I_{A_2}\ket{\phi} $'s for $ \ket{\phi} \in \mathcal{U} $  are mutually orthogonal, for every order pairs $\{\ket{\psi},\ket{\phi}\}, \ket{\phi} \neq \ket{\psi} \in \mathcal{U}_3^{0,2} $ and for every outcome $ \alpha $, $ \mel{\psi}{E_{\alpha}^{AB} \otimes I_{3}^{C}}{\phi} = 0$.	
Now considering the order pairs $ \{\ket{\psi},\ket{\phi}\} $  for $\ket{\psi} \in \{	|\downarrow \kappa^{(1)}_{2,0}\rangle^{-},	|\uparrow \kappa^{(1)}_{2,0}\rangle^{-},	|\downarrow \kappa^{(0)}_{0,2}\rangle^{-},	|\uparrow \kappa^{(0)}_{0,2}\rangle^{-}\}$ and $\ket{\phi}\in\{|\downarrow\kappa^{(1)}_{0,2}\rangle^{-},	|\uparrow \kappa^{(1)}_{0,2}\rangle^{-},	|\downarrow \kappa^{(0)}_{2,0}\rangle^{-},	|\uparrow \kappa^{(0)}_{2,0}\rangle^{-}\}$, we get
$ \alpha_{ij}=0$ (and hence, $ \alpha_{ji}=0$ ) for $ i=3,2,8,7 $ and $j=5,6,0,1 $ respectively. 
\begin{equation}
	\therefore,~~ E_{\alpha}^{A_0A_1}=
	\begin{bmatrix}
		\alpha_{00} & \alpha_{01} & 0 & 0 & \alpha_{04} & \alpha_{05} & \alpha_{06} & 0 & 0 \\
		\alpha_{10} & \alpha_{11} & 0 & 0 & \alpha_{14} & \alpha_{15} & \alpha_{16} & 0 & 0 \\
		0 & 0 & \alpha_{22} & \alpha_{23} & \alpha_{24} & 0 & 0 & \alpha_{27} & \alpha_{28} \\
		0 & 0 & \alpha_{32} & \alpha_{33} & \alpha_{34} & 0 & 0 & \alpha_{37} &  \alpha_{38} \\
		\alpha_{40} & \alpha_{41} & \alpha_{42} & \alpha_{43} & \alpha_{44} & \alpha_{45} & \alpha_{46} & \alpha_{47} & \alpha_{48} \\
		\alpha_{50} & \alpha_{51} & 0 & 0 & \alpha_{54} & \alpha_{55} & \alpha_{56} & 0 & 0 \\
		\alpha_{60} & \alpha_{61} & 0 & 0 & \alpha_{64} & \alpha_{65} & \alpha_{66} & 0 & 0 \\
		0 & 0 & \alpha_{72} & \alpha_{73} & \alpha_{74} & 0 & 0 & \alpha_{77} & \alpha_{78} \\
		0 & 0 & \alpha_{82} & \alpha_{83} & \alpha_{84} & 0 & 0 & \alpha_{87} & \alpha_{88} 
	\end{bmatrix}
\end{equation}
Now for the order pairs $ \{\ket{\psi},|\uparrow\kappa^{(2)}_{2,0}\rangle^{-}\} $, $$\ket{\psi} \in \{|\downarrow \kappa^{(1)}_{2,0}\rangle^{-},|\uparrow \kappa^{(1)}_{2,0}\rangle^{-},|\downarrow \kappa^{(0)}_{0,2}\rangle^{-},	|\uparrow \kappa^{(0)}_{0,2}\rangle^{-}\}$$ and $ \{|\uparrow\kappa^{(2)}_{0,2}\rangle^{-},\ket{\phi}\} $, $$\ket{\phi} \in \{|\downarrow \kappa^{(1)}_{0,2}\rangle^{-},	|\uparrow \kappa^{(1)}_{0,2}\rangle^{-},	|\downarrow \kappa^{(0)}_{2,0}\rangle^{-},	|\uparrow \kappa^{(0)}_{2,0}\rangle^{-}\}$$, we get $ \alpha_{i4}=0$, $ i=2,3,7,8$ (and hence, $ \alpha_{4i}=0$ ) and $ \alpha_{4j}=0$, $j=6,5,1,0 $ (and hence, $ \alpha_{j4}=0$ ) respectively.	
\begin{equation}
	\therefore E_{\alpha}^{A_0A_1}=
	\begin{bmatrix}
		\alpha_{00} & \alpha_{01} & 0 & 0 & 0 & \alpha_{05} & \alpha_{06} & 0 & 0 \\
		\alpha_{10} & \alpha_{11} & 0 & 0 & 0 & \alpha_{15} & \alpha_{16} & 0 & 0 \\
		0 & 0 & \alpha_{22} & \alpha_{23} & 0 & 0 & 0 & \alpha_{27} & \alpha_{28} \\
		0 & 0 & \alpha_{32} & \alpha_{33} & 0 & 0 & 0 & \alpha_{37} &  \alpha_{38} \\
		0 & 0 & 0 & 0 & \alpha_{44} & 0 & 0 & 0 & 0 \\
		\alpha_{50} & \alpha_{51} & 0 & 0 & 0 & \alpha_{55} & \alpha_{56} & 0 & 0 \\
		\alpha_{60} & \alpha_{61} & 0 & 0 & 0 & \alpha_{65} & \alpha_{66} & 0 & 0 \\
		0 & 0 & \alpha_{72} & \alpha_{73} & 0 & 0 & 0 & \alpha_{77} & \alpha_{78} \\
		0 & 0 & \alpha_{82} & \alpha_{83} & 0 & 0 & 0 & \alpha_{87} & \alpha_{88} 
	\end{bmatrix}
\end{equation}	
By choosing the order pairs $ \{|\downarrow\kappa^{(1)}_{2,0}\rangle^{-},\ket{\phi}\} $, $$\ket{\phi} \in \{|\uparrow\kappa^{(1)}_{2,0}\rangle^{-},|\downarrow\kappa^{(0)}_{0,2}\rangle^{-},|\uparrow\kappa^{(0)}_{0,2}\rangle^{-}\}$$ we get 
\begin{equation}
	\begin{array}{l}
		\alpha_{32}+\alpha_{23}=0, 
		\alpha_{27}+\alpha_{38}=0, 
		\alpha_{37}+\alpha_{28}=0.			
	\end{array}
\label{e5}
\end{equation}
For $\{|\downarrow\kappa^{(1)}_{2,0}\rangle^{-},\ket{S}\}$, we get
\begin{equation}
	\{(\alpha_{22}-\alpha_{33})-(\alpha_{32}-\alpha_{23})\}+\{(\alpha_{27}-\alpha_{38})-(\alpha_{37}-\alpha_{28})\}=0
	\label{e1}
\end{equation}
For $\{|\downarrow\kappa^{(1)}_{2,0}\rangle^{-},|\kappa^{(2)}_{0,2}\rangle^{-}\}$, we get
\begin{equation}
	\{(\alpha_{22}-\alpha_{33})-(\alpha_{32}-\alpha_{23})\}-\{(\alpha_{27}-\alpha_{38})-(\alpha_{37}-\alpha_{28})\}=0
	\label{e2}
\end{equation}
(\ref{e1}) and (\ref{e2}) together provides,
\begin{equation}
	\begin{array}{l}
		(\alpha_{22}-\alpha_{33})-(\alpha_{32}-\alpha_{23})=0,\\
		(\alpha_{27}-\alpha_{38})-(\alpha_{37}-\alpha_{28})=0.			
	\end{array}
\label{e3}
\end{equation}
Now by choosing the order pairs $ \{|\downarrow\kappa^{(1)}_{2,0}\rangle^{-},\ket{\phi}\} $, $\ket{\phi} \in \{|\updownarrow\kappa^{(1)}_{2,0}\rangle^{-},|\updownarrow\kappa^{(0)}_{0,2}\rangle^{-}\}$, we get 
\begin{equation}
	\begin{array}{l}
		(\alpha_{22}-\alpha_{33})+(\alpha_{32}-\alpha_{23})=0,\\
		(\alpha_{27}-\alpha_{38})+(\alpha_{37}-\alpha_{28})=0.			
	\end{array}
\label{e4}
\end{equation}
(\ref{e3}) and (\ref{e4}) togetherly gives
\begin{equation}
	\begin{array}{l}
		(\alpha_{22}-\alpha_{33})=0, 
		(\alpha_{32}-\alpha_{23})=0\\
		(\alpha_{27}-\alpha_{38})=0, 
		(\alpha_{37}-\alpha_{28})=0.			
	\end{array}
\label{e6}
\end{equation}
From (\ref{e5}) and (\ref{e6}), we get $ \alpha_{32}=\alpha_{23}=\alpha_{27}=\alpha_{38}=\alpha_{37}=\alpha_{28}=0$ and $\therefore \alpha_{72}=\alpha_{83}=\alpha_{73}=\alpha_{82}=0$.	
Now for $ \{\ket{\psi},|\downarrow\kappa^{(1)}_{0,2}\rangle^{-}\} $, $\ket{\psi} \in \{|\downarrow\kappa^{(0)}_{2,0}\rangle^{-},|\uparrow\kappa^{(0)}_{2,0}\rangle^{-}\}$, we get 
\begin{equation}
	\begin{array}{l}
		\alpha_{05}+\alpha_{16}=0, 
		\alpha_{15}+\alpha_{06}=0			
	\end{array}
\label{e7}
\end{equation}	
Now by considering $\{|\downarrow\kappa^{(0)}_{2,0}\rangle^{-},|\updownarrow\kappa^{(1)}_{0,2}\rangle^{-}\}$ we get 
\begin{equation}
	(\alpha_{05}-\alpha_{16})+(\alpha_{15}-\alpha_{06})=0
	\label{e8}
\end{equation} 
and by choosing 
$\{|\downarrow\kappa^{(0)}_{2,0}\rangle^{-},|\kappa^{(0)}_{0,2}\rangle^{-}\}$ we get 
\begin{equation}
	(\alpha_{05}-\alpha_{16})-(\alpha_{15}-\alpha_{06})=0
	\label{e9}
\end{equation}
(\ref{e8}) and (\ref{e9}) gives together
\begin{equation}
	\begin{array}{l}
		(\alpha_{05}-\alpha_{16})=0, 
		(\alpha_{15}-\alpha_{06})=0.			
	\end{array}
\label{e10}
\end{equation}
From (\ref{e7}) and (\ref{e10}), we get $ \alpha_{05}=\alpha_{16}=\alpha_{15}=\alpha_{06}=0$ and $\therefore \alpha_{50}=\alpha_{61}=\alpha_{51}=\alpha_{60}=0$.	
Choosing order pairs $\{|\downarrow\kappa^{(0)}_{2,0}\rangle^{-},|\uparrow\kappa^{(0)}_{2,0}\rangle^{-}\},\{|\downarrow\kappa^{(1)}_{0,2}\rangle^{-},|\uparrow\kappa^{(1)}_{0,2}\rangle^{-}\}$ and $\{|\downarrow\kappa^{(0)}_{0,2}\rangle^{-},|\uparrow\kappa^{(0)}_{0,2}\rangle^{-}\}$ we get
\begin{equation}
	\begin{array}{l}
		(\alpha_{10}+\alpha_{01})=0, 
		(\alpha_{65}+\alpha_{56})=0, 
		(\alpha_{78}+\alpha_{87})=0.		
	\end{array}
\label{e14}
\end{equation}	
Now for the pairs $\{|\downarrow\kappa^{(0)}_{2,0}\rangle^{-},|\updownarrow\kappa^{(0)}_{2,0}\rangle^{-}\},\{|\downarrow\kappa^{(1)}_{0,2}\rangle^{-},|\updownarrow\kappa^{(1)}_{0,2}\rangle^{-}\}$ and $\{|\downarrow\kappa^{(0)}_{0,2}\rangle^{-},\ket{S}\}$ we get
\begin{equation}
	\begin{array}{l}
		(\alpha_{00}-\alpha_{11})+(\alpha_{10}-\alpha_{01})=0\\
		(\alpha_{55}-\alpha_{66})+(\alpha_{65}-\alpha_{56})=0\\
		(\alpha_{77}-\alpha_{88})+(\alpha_{78}-\alpha_{87})=0
		\label{e11}			
	\end{array}
\end{equation}
and for the pairs $\{|\updownarrow\kappa^{(0)}_{2,0}\rangle^{-},|\downarrow\kappa^{(0)}_{2,0}\rangle^{-}\},\{|\updownarrow\kappa^{(1)}_{0,2}\rangle^{-},|\downarrow\kappa^{(1)}_{0,2}\rangle^{-}\}$ and $\{\ket{S},|\downarrow\kappa^{(0)}_{0,2}\rangle^{-}\}$ we get
\begin{equation}
	\begin{array}{l}
		(\alpha_{00}-\alpha_{11})-(\alpha_{10}-\alpha_{01})=0\\
		(\alpha_{55}-\alpha_{66})-(\alpha_{65}-\alpha_{56})=0\\
		(\alpha_{77}-\alpha_{88})-(\alpha_{78}-\alpha_{87})=0
		\label{e12}			
	\end{array}
\end{equation}
(\ref{e11}) and (\ref{e12}) gives 
\begin{equation}
	\begin{array}{l}
		(\alpha_{00}-\alpha_{11})=0, 
		(\alpha_{10}-\alpha_{01})=0\\
		(\alpha_{55}-\alpha_{66})=0, 
		(\alpha_{65}-\alpha_{56})=0\\
		(\alpha_{77}-\alpha_{88})=0, 
		(\alpha_{78}-\alpha_{87})=0.			
	\end{array}
\label{e13}
\end{equation}
From (\ref{e14}) and (\ref{e13}), we get $ \alpha_{10}=\alpha_{01}=\alpha_{65}=\alpha_{56}=\alpha_{78}=\alpha_{ 87}=0$.
\begin{equation}
	\therefore E_{\alpha}^{A_0A_1}=
	\begin{bmatrix}
		\alpha_{00} & 0 & \cdots & 0 \\
		0 & \alpha_{11} & \cdots & 0 \\
		\vdots & \vdots & \ddots & \vdots \\
		0 & 0 & \cdots & \alpha_{88}
	\end{bmatrix}
\end{equation}
is now a diagonal matrix where $ \alpha_{00}=\alpha_{11},\alpha_{22}=\alpha_{33},\alpha_{55}=\alpha_{66},\alpha_{77}=\alpha_{88}$.
Now considering the pairs $\{|\downarrow\kappa^{(2)}_{0,2}\rangle^{-},|\kappa^{(1)}_{0,2}\rangle^{-}\}$,$\{|\uparrow\kappa^{(2)}_{2,0}\rangle^{-},|\updownarrow\kappa^{(2)}_{2,0}\rangle^{-}\}$ and $\{|\downarrow\kappa^{(2)}_{2,0}\rangle^{-},|\updownarrow\kappa^{(2)}_{2,0}\rangle^{-}\}$, 
we get
\begin{equation}
	\begin{array}{l}
		\alpha_{22}=\alpha_{77}, 
		\alpha_{44}=\alpha_{55}, 
		\alpha_{11}=\alpha_{66}.			
	\end{array}
\end{equation} 
Now by choosing the pair $\{|\kappa^{(1)}_{0,2}\rangle^{-},\ket{S}\}$
we get
\begin{equation}
	\begin{array}{l}
		2(\alpha_{00}+\alpha_{11})-\alpha_{44}=\alpha_{22}+\alpha_{33}+\alpha_{77}\\
		\Rightarrow 2(\alpha_{00}+\alpha_{00})-\alpha_{00}=\alpha_{22}+\alpha_{22}+\alpha_{22}\\
		\Rightarrow 3(\alpha_{00})=3(\alpha_{22})\\
		\Rightarrow \alpha_{00}=\alpha_{22}			
	\end{array}
\end{equation}
Eventually we get $\alpha_{00}=\alpha_{11}=\alpha_{22}=\alpha_{33}=\alpha_{44}=\alpha_{55}=\alpha_{66}=\alpha_{77}=\alpha_{88}$.$\blacksquare$\\

\end{document}